\documentclass[english,aps,prl,amsmath,twocolumn,preprintnumber,superscriptaddress,notitlepage,floatfix]{revtex4-2}

\usepackage[T1]{fontenc}
\makeatletter
\usepackage[normalem]{ulem}
\usepackage{amsfonts}
\usepackage{graphicx}% Include figure files
\graphicspath{ {figures/}{figures/eps/}{figures/pdf/} }% specify the path where figures are located
\usepackage{dcolumn}% Align table columns on decimal point
\usepackage{bm}% bold math
\usepackage{amssymb} % for \hbar and other math symbols

\usepackage{hyperref}% add hypertext capabilities
 \hypersetup{
    colorlinks=true,   % false: boxed links; true: colored links
    linkcolor=blue,    % color of internal links (e.g., sections)
    citecolor=blue,   % color of links to bibliography
    urlcolor=cyan      % color of external links
}
\usepackage{nomencl}                    % produces a nomenclature
\usepackage{float}                      % figure floats
\usepackage{fancyhdr}                   % fancy headers and footers
\usepackage{url}                        % nicely format url breaks
\usepackage[inactive]{srcltx}		 	% necessary to use forward and inverse searching in DVI
\usepackage{relsize}                    % font sizing hierarchy
\usepackage{booktabs}                   % professional looking tables
\usepackage{mathrsfs}                   % additional math scripts
\usepackage[titletoc]{appendix}			% format appendix correctly
\usepackage{comment}
\usepackage{dsfont}
\usepackage{amsmath}
\usepackage{mathtools}
\usepackage{mathptmx}
\usepackage{newtxmath}
\usepackage{amssymb}
\usepackage{array}
\usepackage[utf8]{inputenc}
\usepackage{color}
\usepackage{bm}
\usepackage{subcaption}
\usepackage{wrapfig}
\usepackage[percent]{overpic} % for overlaying images
\usepackage{braket}
\usepackage{feynmp-auto}
\usepackage[compat=1.1.0]{tikz-feynman}
\tikzfeynmanset{warn luatex=false} % Suppress warnings related to LuaTeX
\usepackage{siunitx}
\usepackage{qcircuit}
\DeclarePairedDelimiterXPP\BigO[1]{\mathcal{O}}{\lparen}{\rparen}{}{#1}
\DeclareUnicodeCharacter{2212}{-}
\DeclareUnicodeCharacter{2265}{\ensuremath{\geq}}
\usepackage{tikz}

\renewcommand{\hbar}{\mathchar'26\mkern-9mu h}

\newcommand{\sr}[1]{\textcolor{blue}{#1}} 

\begin{document}

% \preprint{APS/123-QED}

\title{Two-beam Multiparticle Many-body simulations of \\Inhomogeneous FFI}% Force line breaks with \\
\author{Zoha Laraib}
\email{zlaraib@vols.utk.edu}
\affiliation{Department of Physics and Astronomy, The University of Tennessee, Knoxville, Tennessee 37996, USA}
\affiliation{National Center for Computational Sciences, Oak Ridge National Laboratory, Oak Ridge, Tennessee 37830, USA}
\author{Sherwood Richers}
\affiliation{Department of Physics and Astronomy, The University of Tennessee, Knoxville, Tennessee 37996, USA}

\date{\today}% It is always \today, today,

\begin{abstract}
Neutrino flavor evolution in dense astrophysical environments is inherently nonlinear and sensitive to many-body (MB) quantum effects beyond the mean-field (MF) approximation. Existing MB studies are constrained by small system sizes, closed boundaries, and highly idealized symmetry assumptions. We present a unified tensor-network framework that enables simulations of inhomogeneous and anisotropic flavor evolution under conditions relevant to core-collapse supernovae and neutron-star mergers. Within this framework, we examine the effects of inhomogeneity, boundary conditions, and convergence with resolution for multiple neutrino distributions, allowing direct comparison of these setups under one consistent formulation. In our simulations, many-body systems equilibrate earlier than their mean-field counterparts while approaching similar final flavor states. Enlarging the interaction region allows open boundaries to reproduce closed-system behavior, but only when the beams begin superimposed and interact continuously. By contrast, initially separated configurations develop entanglement more slowly, interact over longer times, and equilibrate to a flavor content that differs from that obtained from initially superimposed calculations.
\end{abstract}

\maketitle
\section{Introduction}\label{sec:introduction}

Astrophysical environments such as core-collapse supernovae (CCSNe) and neutron star mergers (NSMs) serve as natural laboratories for exploring dense matter, strong gravity, and neutrino quantum kinetics~\cite{janka_ExplosionMechanismsCoreCollapse_2012,mezzacappa_CoreCollapseSupernova_2023,foucart2023neutrino,burrows2021core}. In these systems, neutrinos play a dominant role in energy and lepton transport, determining the neutron-to-proton ratio and shaping nucleosynthesis in the ejecta~\cite{Qian1993,wang2022neutrinos,fischer2024neutrinos}. Consequently, accurate modeling of neutrino flavor evolution is crucial for interpreting future neutrino and gravitational-wave observations from such events.

In vacuum and ordinary matter, neutrino oscillations are driven by the PMNS mixing matrix and forward scattering on electrons~\cite{wolfenstein2018neutrino,mikheyev1989resonant}. In dense astrophysical environments, neutrino self-interactions give rise to nonlinear collective phenomena such as spectral swaps, bipolar oscillations, and fast flavor instabilities (FFIs)~\cite{pantaleone1992neutrino,duan2010collective,tamborra2021new,capozzi2022neutrino,richers_FastFlavorTransformations_2022, johns2025neutrino, volpe2024neutrinos}. These FFIs originate from angular crossings in the neutrino angular distributions deep below the shock and can develop on nanosecond timescales~\cite{sawyer2005speed,banerjee2011linearized,dasgupta2017fast}.

FFIs in quantum kinetic theory is largely observed in the Mean-field (MF) approaches ~\cite{fiorillo2024inhomogeneous, Shalgar2021breaking} that approximate the many-body (MB) density matrix by factorization, thereby neglecting quantum entanglement, correlations, and decoherence effects ~\cite{patwardhan2022many,balantekin2023quantum}. Nevertheless, these MF approaches remain valuable even in homogeneous settings, where they provide insight into phenomena such as spontaneous symmetry breaking ~\cite{raffelt2013neutrino,hansen2014chaotic,chakraborty2016self,duan2015flavor,abbar2015flavor,mangano2014damping,mirizzi2015self}. However, despite capturing coherent flavor oscillations effectively, these MF treatments ultimately break down in regimes involving finite-size systems or strong correlations induced by neutrino self-interactions.

The MB approaches preserve quantum correlations by representing neutrinos as interacting spins~\cite{bell2003speed,friedland2003many,balantekin2006neutrino,patwardhan_ManybodyCollectiveNeutrino_2023,roggero2021entanglement}. However, these methods remain severely constrained by the exponential scaling of the Hilbert space, limiting simulations to 
 $\mathcal{O}(10$–$1000)$ particles even with tensor-network techniques~\cite{roggero2021dynamical,roggero2021entanglement,roggero2022entanglement,martin2023equilibration,illa2023multi,bhaskar2024timescales}. To stay computationally feasible, most MB studies impose strong idealizations— isotropy~\cite{cervia2022collective, roggero2021entanglement,roggero2021dynamical}, fixed beam geometries~\cite{roggero2021dynamical,roggero2022entanglement}, or periodic boundaries that force neutrinos to re-encounter each other and artificially enhance coherence and entanglement~\cite{martin2023equilibration,shalgar2023we}. Other limitations include using finite neutrino interaction length/times ~\cite{shalgar2023we,kost2024once} as a proxy for interacting plane waves or neutrino beams, neglect momentum-changing (non-forward) scattering~\cite{johns2023neutrino,cirigliano2024neutrino}, and Pauli blocking~\cite{goimil2025pauli}. An alternative framework for describing quantum correlations in collective oscillations is based on the Bogoliubov-BornGreen-Kirkwood-Yvon (BBGKY) hierarchy ~\cite{volpe2013extended}. In simplified homogeneous two-beam systems, tensor network methods have enabled tractable MB simulations of a larger number of sites~\cite{roggero2021dynamical,Cervia_2022,Siwach_2023}. However, recent work demonstrates that introducing even modest spatial variations can significantly impact the development of FFIs—even in large systems~\cite{laraib2025many}. This raises the question of whether relaxing these idealizations fundamentally changes MB dynamics and whether such effects persist under realistic astrophysical conditions.

% \sr{[[This is redundant with the following paragraph, isn't it?]]} In this work, we systematically relax these symmetry and confinement assumptions to reveal how geometry and spatial overlap shape many-body flavor evolution. Specifically, we: (1) evaluate open boundary conditions that emulate realistic streaming behavior near the neutrinosphere, enabling causal, single-pass interactions, and compare against closed boundaries; (2) investigate superimposed and spatially separated beam configurations to disentangle geometric and overlap effects, capturing distinct interaction timescales and coherence properties relevant to neutrino crossings near the neutrinosphere; (3) demonstrate density-scaling analyses aimed at disentangling genuine many-body correlations from finite-size artifacts by identifying configurations that preserve physical entanglement and approach thermodynamic-limit behavior; and (4) analyse asymmetric $\nu_e$–$\nu_\mu$ systems alongside flavor-symmetric cases to reveal how lepton asymmetry affects the MF instability growth, and how it compares to MB equilibration, and many-body coherence. Collectively, our results bridge the MF and MB regimes, establishing a robust foundation for realistic modeling of neutrino flavor evolution in CCSNe and NSMs. 

We organize the paper as follows. In Section~\ref{sec:Methods}, we describe our tensor-network framework to model a forward-scattering Hamiltonian with spin–spin couplings in neutrino–neutrino interactions, including anisotropic angular structure and spatial inhomogeneity, while section~\ref{sec:results} describes our two-beam initialization of counter-propagating $\nu_e$ and $\nu_\mu$ streams. We first examine how homogeneous (Section~\ref{sec:Homogeneous FFI}) and inhomogeneous (Section~\ref{sec:results_inhomo}) spatial configurations influence the onset and equilibration of flavor transformation. In Section~\ref{sec:results_inhomo} we systematically step through the transition from MF to MB by varying the bond dimension, and investigate the impact of open and closed boundary conditions, initial conditions of superimposed vs spatially separated neutrinos, and asymmetric numbers of neutrinos of different flavors in the initial conditions. This framework helps us extend the MB literature by providing the first consistent, physically motivated study of homogeneity, asymmetry, boundary conditions, spatial structure, all under the same structure.

\section{Methods}\label{sec:Methods}
We start by looking at the solution of the time-dependent many-body Schrodinger equation for a general quantum state
\begin{equation}
    |\psi\rangle = \sum_{\sigma_1,...,\sigma_{N_\mathrm{sites}}} c_{\sigma_1,...,\sigma_{N_\mathrm{sites}}} |\sigma_1,...,\sigma_{N_\mathrm{sites}}\rangle
    \label{eq:2.2 TN}
\end{equation}
consisting of $N_\mathrm{sites}$ local, spin-like degrees of freedom $\sigma_i \in \{\uparrow,\downarrow\}$, where $|\uparrow \rangle $ represents an electron flavor state and $|\downarrow \rangle$ represents a muon flavor state. We assume two flavors throughout this work. This state is fully defined by the rank $N_\mathrm{sites}$ tensor with components $c_{\sigma_1,...,\sigma_{N_\mathrm{sites}}} \in \mathbb{C}$. In order to efficiently evolve states with small entanglement entropy, we use the ITensor and ITensorMPS \cite{fishman2022itensor,fishman2022codebase} libraries in Julia. These libraries decompose the tensor $c_{\sigma_1,...,\sigma_{N_\mathrm{site}}}$ into a product of smaller tensors \cite{vidal2003efficient,vidal2004efficient,vidal2007classical,verstraete2008matrix} and use truncated singular-value decompositions SVDs) to compress the quantum state \cite{roggero2021entanglement, paeckel2019time,schollwock2011density}. The bond dimension (BD) describes the number of singular values accounted for in the connection between two sites. While the maximum bond dimension of a MPS is $2^{N_\mathrm{sites}/2}$, the BD can be chosen to be any smaller value, although the truncation of the SVD may destroy important information in the quantum state \cite{roggero2021entanglement,vidal2003efficient}. We investigate the BD required for accurate calculations in Section~\ref{sec:results}.

We evolve the many-body quantum state using a time-dependent Hamiltonian
\begin{equation}
    \frac{\partial \psi}{\partial t} = -i H[Z(t)] \psi \label{eq:Schrodinger eq}
\end{equation}
where $Z(t) = \{\vec{z}_i\}(t)$ is the set of time-varying positions of each site indexed by $i$. We assume that each site represents a cubic volume of side length $w$ containing $N$ neutrinos, each of which has momentum $\vec{p}_i$. The position of each site evolves as $d\vec{z}_i/dt = c\hat{p}$, where $c$ (without indices) is the speed of light and $\hat{p}=\vec{p}_i/|\vec{p}_i|$. The Hamiltonian can be split into constant and time-independent terms
\begin{equation}
    H[Z(t)] = H_\mathrm{vac} + H_\mathrm{SI}[Z(t)]\,\,. \label{eq:collective_osc}
\end{equation}
The constant vacuum Hamiltonian is
\begin{equation}
\label{eq:H_vac}
    H_\mathrm{vac} = \sum_i \frac{\Delta m^2}{2|\vec{p}_i|} \vec{B}\cdot\vec{\sigma}_i\,\,,
\end{equation}
where \(\Delta m^2 = m_2^2-m_1^2\) is the mass-splitting and \(\vec{\sigma}_i\) denotes the vector of Pauli matrices acting on site \(i\). Matter effects can be treated in this framework via an effective mass splitting and mixing angle in the presence of a dense background, but we will continue referring to this as the vacuum term for simplicity. The unit vector \(\vec{B}\) encodes the mixing angle \(\theta\) as \(\vec{B} = ( \sin(2\theta), 0, -\cos(2\theta))\), corresponding to the normal hierarchy when $\theta$ is small. 

The neutrino self-interaction Hamiltonian is given by  
\begin{equation}
%H_\mathrm{SI}[Z(t)] = \frac{\sqrt{2}G_F}{2} \sum_{i<j} \frac{N_i+N_j}{w^3}\vec{\sigma_i} \cdot \vec{\sigma_j} J_{ij} \mathcal{S}(\xi_{ij})\,\,,
H_\mathrm{SI}[Z(t)] = \sqrt{2}G_Fn \sum_{i<j}\vec{\sigma_i} \cdot \vec{\sigma_j} J_{ij} \mathcal{S}(\xi_{ij})\,\,,
\label{eq:self-int}
\end{equation}
where the subscripts i and j are site indices and $n=N/w^3$ is the neutrino number density at each site. The directional dependence of the neutrino-neutrino interaction is encoded in the geometric factor
\(J_{ij}=(1 - \hat{p}_i \cdot \hat{p}_j) \)  where \( \hat{p}_j = \vec{p}_j / |\vec{p}_j| \) is the direction of the \(j\)th neutrino. However, we can also set \(J_{ij} = 1\) when comparing to prior literature that assumes isotropy.

In addition, we account for spatial inhomogeneity by introducing a shape function \( \mathcal{S}(\xi_{ij}) \), where the input \(\xi_{ij} = (z_i - z_j)/ w \) is the distance between two sites normalized by the size of the spatial extent of the site. This function enforces the local nature of the neutrino-neutrino coupling, only allowing sites to interact if they are at the same location. We note that this function does not represent a physical interaction strength, but rather serves as a numerical tool that enforces true interaction locality in the limit of zero site size $w$. Constructed in this way, any local function that integrates to a value of 1 is a valid numerical tool, though we choose: 
    % \mathcal{S}_\mathrm{flat-top}(\xi)&=\Theta(0.5-|\xi|)\,\,, \\
\begin{equation}
\begin{aligned}
    \mathcal{S}(\xi_{ij}) &= (1-|\xi_{ij}|)\Theta(1-|\xi_{ij}|)\,\,,
\end{aligned}
\end{equation}
where $\Theta$ is the Heaviside step function. 
When assuming a homogeneous environment to connect with prior literature we simply set \(\mathcal{S}(\xi_{ij})=1\). This treatment of inhomogeneity is not dissimilar from that of \cite{shalgar2023we}, though our parameterization allows for a straightforward separation of computational and physical parameters.

\subsection{Boundary conditions}

We use two types of boundary conditions: Open and Closed (i.e., periodic). Open boundary conditions allow sites to drift infinitely far, eventually allowing sites to stop interacting altogether when they become distant. 

Closed boundary conditions approximately simulate environments that are homogeneous on scales larger than the domain size $L$. To do so, we identify the position $z$ with the positions $z\pm L$. This requires that particle positions wrap around the domain, so site positions are modified as $z_i \leftarrow (z_i \bmod L)$ at the end of each timestep. In addition, pairwise interactions must wrap around as well. To do this,
the minimal separation between particle pairs i and j must reflect the shortest distance considering the periodicity of the domain. Hence, the argument to the shape function is modified as: 
\begin{equation}
\xi \leftarrow 
\begin{cases}
\xi & |z_i-z_j| \leq \phantom{-}L/2 \\[6pt]
\xi - L/w &  \phantom{|}z_i - z_j\phantom{|} > \phantom{-}L/2 \\[6pt]
\xi + L/w &  \phantom{|}z_i - z_j\phantom{|} < -L/2
\end{cases}\,\,.
\end{equation}

\subsection{Time evolution}
 
We use time-evolving block decimation (TEBD) to approximate the time evolution operator \cite{vidal2004efficient} using a second-order Trotter-Suzuki decomposition. That is, for a time step of size $\delta$ the exact time evolution operator (assuming geometric and shape terms are constant within a timestep) is
\begin{equation}
     \hat{U}_\mathrm{exact}(\delta) = e^{-i\hat{H}\delta}\,\,,
\end{equation}
where $\hat{H}$ represents the full many-body Hamiltonian. This can be broken down into a series of successive one- and two-site evolution operators using a second-order Trotter-Suzuki decomposition \cite{paeckel2019time}:

\begin{equation}
\hat{U}_{\text{TEBD2}}(\delta) = \prod_{\alpha=1}^{\mathcal{N}_\mathrm{gates}} e^{-i \hat{H}_\alpha \delta /2} \prod^{1}_{\alpha=\mathcal{N}_\mathrm{gates}} e^{-i \hat{H}_\alpha \delta /2}\,\,.
\end{equation}
This scheme approximates the full Hamiltonian with error scaling as 
% \begin{equation}
% \hat{U}_{\text{TEBD2}}(\delta) = e^{-i\hat{H}_{odd}\delta/2}e^{-i\hat{H}_{even}\delta}e^{-i\hat{H}_{odd}\delta/2}. \label{eq:tebd I} 
% \end{equation}
$\hat{U}_{\text{exact}}(\delta) = \hat{U}_{\text{TEBD2}}(\delta) + O(\delta^3)$. The $H_\alpha$ are the collection of terms in the sums in Equations \ref{eq:H_vac} and \ref{eq:self-int}.

In addition, we reorder the sites after each timestep such that the spatial ordering of the sites is reflected in the ordering in the MPS. This keeps interactions local and increases the accuracy of finite-BD simulations.

\subsection{Reduced states and One-particle Entanglement}\label{sec:one-particle Entanglement}

Measures of quantum entanglement are useful probes of the many-body nature of solutions that cannot be represented by the mean-field approximation. We use the one-particle Von Neumann entanglement entropy given in Refs.~\cite{cervia2019entanglement,patwardhan2021spectral} via the relation
\begin{equation}
S_i = -\frac{1 - P_i}{2} \log\left( \frac{1 - P_i}{2} \right) - \frac{1 + P_i}{2} \log\left( \frac{1 + P_i}{2} \right)\,\,, \label{eq:entropy}
\end{equation}
where $\vec{P}_i=\langle\vec{\sigma}_i\rangle$ is the one-body reduced polarization vector of site $i$ and $P_i=|\vec{P_i}|$ is the magnitude of the polarization vector. In this context, \( P \) serves as an indicator of entanglement: \( P = 1 \) corresponds to a pure state, while \( P < 1 \) signifies that the reduced density matrix is mixed, indicating the presence of entanglement between the site and the rest of the system. The mean-field assumption requires that $S_i=0$ (i.e., $P_i=1$) at every site, while the full many-body description of neutrino cloud dynamics allows for \( P < 1 \) and $0<S<1$.

From the polarization vectors, we can also construct one-site reduced density matrices as
\begin{equation}
\rho_i = \begin{pmatrix}
\rho_{ee,i} & \rho_{e \mu,i} \\
\rho_{\mu e,i} & \rho_{xx,i}
\end{pmatrix} =
\frac{1}{2}
\begin{pmatrix}
1 + P_{z,i} & P_{x,i} + iP_{y,i} \\
P_{x,i} - iP_{y,i} & 1 - P_{z,i} \\
\end{pmatrix}\,\,,\label{eq:reduced_density_II}
\end{equation}
where the diagonals of the density matrix, gives occupation number information of electron and muon neutrino respectively. The off-diagonals gives information about the flavor coherence at each site.

\section{Results}\label{sec:results}
% We use the two-beam geometry as it encapsulates self-induced neutrino flavor conversion that influences energy deposition beyond the decoupling region, neutrino-driven nucleosynthesis, and detection opportunities of the neutrino signal from the next nearby supernova or the diffuse supernova neutrino flux from past core-collapse events \cite{janka2012explosion,scholberg2012supernova, mirizzi2016supernova}.

We restrict our model to the simplest possible configuration of spatial and momentum - a one-dimensional system akin to two directly opposing $\nu_e$ - $\nu_{\mu}$ neutrino colliding  beams - to isolate essential characteristics common to more complex angular distributions, including the development of fast instabilities in the mean-field limit. This geometry highlights the rapid onset of flavor instabilities driven primarily by the neutrino-neutrino interaction energy scale, $\mu = \sqrt{2} G_F n_\nu$. This simplified scenario affords the considerable advantage of yielding fully analytical solutions to the linear regime in the mean field limit\cite{chakraborty2016self2}, thereby providing clear physical insights into the mechanism of rapid flavor conversion without the complications inherent in numerically intensive multi-angle treatments. 

 % Our findings draw on mean-field analytical and simulation analyses from prior studies of inhomogeneous media demonstrating fast instability, such as the ELN crossing sufficient and necessary condition for fast instability \cite{morinaga2022fast} and Particle-in-Cell simulations \cite{richers_ParticleincellSimulationNeutrino_2021}, and extend this analysis for a complete many-body treatment.

%\begin{table*}[]
%    \centering
%    \begin{tabular}{lcccccc}
%    Name & Shape Function & Domain (cm) & Boundaries & $N_{\nu_e}/N_{\nu_\mu}$ & $N_\mathrm{sites}^\mathrm{each\,\,flavor}$\\\hline
%        Homogeneous & Homogeneous & ($-\infty,\infty$) & Periodic & 1 & 1\\
%        Fiducial & Triangular  & (-1,1] & Periodic & 1 & 10 \\
%        Symmetric & Triangular & (-1,1] & Periodic & 1 & 7\\
%        Symmetric $\mathcal{S}=1$ & Homogeneous & (-1,1] & Periodic & 1 & 7\\
%        Asymmetric & Triangular& (-1,1] & Periodic & 2 & 7\\
%        Asymmetric $\mathcal{S}=1$ & Homogeneous & (-1,1] & Periodic & 2 & 7\\
%        Open Superimposed \\
%        Open Separated \\
%    \end{tabular}
%    \caption{Caption}
%    \label{tab:placeholder}
%\end{table*}

\subsection{Homogeneous two-beam FFI}\label{sec:Homogeneous FFI}
Before diving into new results, we first visit the homogeneous 2-beam FFI. Both the mean-field and many-body results are well known, and our goal here is to reproduce them using the full machinery of CCNO. 

This test considers a two-beam model with neutrino masses of $m_1 = 0.008596511$ eV and $m_2 = 0$ eV (i.e. inverted mass hierarchy) and a mixing angle of $\theta_{12} = 1.74532925 \times 10^{-8}$. We only initialize two particles and use the homogeneous shape function ($\mathcal{S}=1$). One particle (containing initially only electron neutrinos) moves in $-\hat{z}$ direction while the other particle (containing initially only muon neutrinos) moves in $+\hat{z}$ direction (i.e., $ |\Psi_0\rangle = |\uparrow \downarrow \rangle$, where the first arrow indicates the state of the left moving particle and the second arrow indicates the state of the right-moving particle). Both particles are assumed to have an energy of 50 MeV. The number of neutrinos in each particle is set such that $n  = (m_2^2 - m_1^2)c^4 / 4 \sqrt{2} G_F |\vec{p}| = 2.92 \times 10^{24} \mathrm{cm}^{-3}$ in order to excite an unstable homogeneous mode.

\begin{figure}
    \centering
    \includegraphics[width=\linewidth]{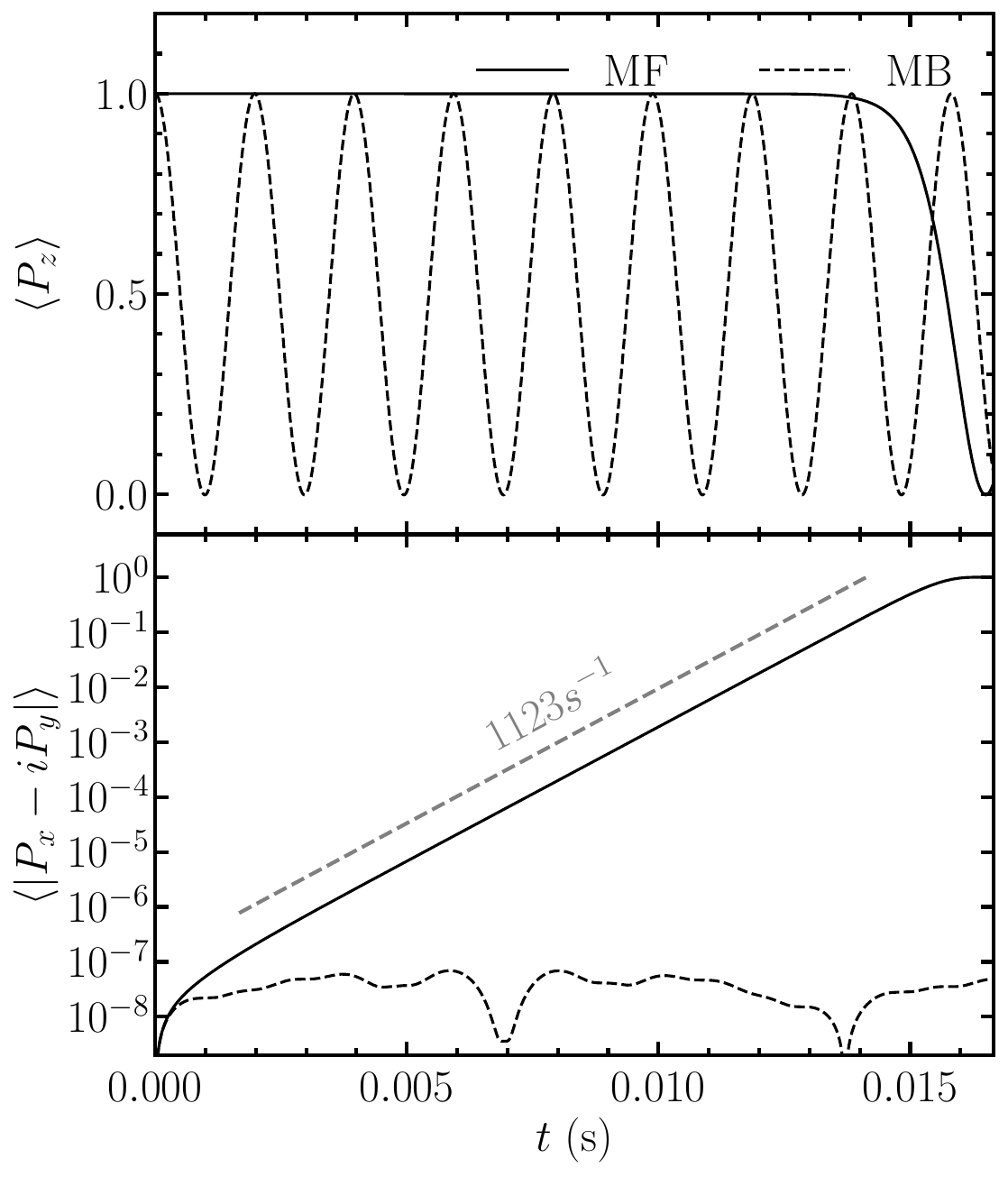}

    \caption{Homogeneous two-flavor two-beam fast flavor instability prepared in an initially half muon and half electron flavor state. The upper panel shows occupation number of electron neutrinos in a mean-field (solid) vs full many-body picture (dashed). The lower panel shows the quantum coherence of electron and muon flavor states showing that the instability grows exponentially for the mean field case but does not for many-body. This growth rate of the flavor  coherence matches the prediction from linear stability theory well. In contrast, many-body entanglement could erase homogeneous FFI.}
    \label{fig:Richers_Homo}
\end{figure}

Figure~\ref{fig:Richers_Homo} shows that we successfully reproduce the homogeneous FFI in the mean-field (i.e., BD=1) limit. The solid curves show the evolution of $\langle P_z \rangle$ (top panel) and $\langle|P_x - iP_y|\rangle$ (bottom panel) of the first particle. Linear stability analysis predicts an exponential increase in the flavor coherence with a growth rate of \( \mathrm{Im}(\Omega) = (m_2^2 - m_1^2)c^4/(2 \hslash h \nu) = 1123 \text{ s}^{-1} \) \cite{chakraborty2016self2, richers2021particle}. The fitted growth rates between \( t = 8.4 \) ms and \( 11.7 \) ms match the theoretical rate with a relative error of \( 1.67\times10^{-3} \). 

We then simulate the same setup, allowing for entanglement by setting BD=2 (the maximum bond dimension for this simple setup) while leaving other parameters unchanged. The dashed curves in Fig.~\ref{fig:Richers_Homo} show that the evolution deviates significantly from the mean-field case: the flavor coherence, which grows exponentially under fast flavor instability (FFI) in the mean-field limit, remains small in the many-body treatment, indicating a qualitatively distinct dynamical behavior driven by entanglement. This behavior is consistent with prior studies~\cite{martin2022classical, martin2023many, balantekin2022quantum, balantekin2023quantum, cervia2019entanglement, cervia2022collective, patwardhan2021spectral, patwardhan2022many, rrapaj2020exact, xiong2022many}, which demonstrate that many-body entanglement can qualitatively alter collective flavor transformation - particularly in small systems where quantum correlations dominate over mean-field-like coherent behavior.

\subsection{Inhomogeneous two-beam FFI}\label{sec:results_inhomo}

We perform simulations of the inhomogeneous two-beam two-flavor fast flavor instability to include the spatial advection of neutrinos using the same neutrino masses, mixing angle, and energy as specified in Section~(\ref{sec:Homogeneous FFI}). In order to engineer a mean-field FFI with a fastest growing mode with wavelength $\lambda=1\,\mathrm{cm}$, we set the neutrino and antineutrino number densities as:
\( n  = 4.89 \times 10^{32} \, \text{cm}^{-3}
\) 
% = \frac{(m_2^2 - m_1^2)c^4/(2h\nu) + \hslash c}{2\sqrt{2}G_F}
(see Equation 2.7 in \cite{chakraborty2016self2}). We place a train of electron neutrinos and a muon neutrinos with propagation directions identical to those in Section~\ref{sec:Homogeneous FFI}. The particles are distributed evenly over space with an electron and a muon neutrino initially collocated at each location with spacing $\Delta z=2\times 1\,\mathrm{cm} / N_{\rm{sites}}$. The particle interactions are determined by a shape function of width $w=\Delta z$, and follow advection in a periodic box of size $L=1\,\mathrm{cm}$. 

We introduce a small perturbation on our initial conditions using one-body potentials that slightly displaces the polarization vector of each neutrino from the \(z\)-axis. This perturbation has a sinusoidal spatial dependence that matches the wavelength of the fastest-growing mode in the mean-field linear stability analysis. That is, we isolate the fastest-growing mode by initializing the off-diagonal density matrix elements as \(\text{Re}(\rho_{e\mu,i}) = \pm10^{-6} \sin(kz_{i})\), with positively perturbed electron neutrinos and negatively perturbed muon neutrinos, where \(z_{i}\) represents the position of the \(i\)-th particle. Its role is to seed the fast flavor instability, initiating flavor evolution in a system with no initial flavor inhomogeneity otherwise. A larger perturbation amplitude shortens the time to saturation of the instability and onset of nonlinear evolution, after which collective neutrino interactions redistribute flavor coherence across momentum modes, driving the system toward depolarization subject to ELN conservation~\cite{bhattacharyya2021fast, izaguirre2017fast, chakraborty2016self, chakraborty2016self2, sawyer2016neutrino}
Here, the growth rate of the instability is primarily set by the local neutrino density, and not by vacuum mixing, which means flavor transformation can still occur in the absence of intrinsic mixing if seeded appropriately~\cite{sawyer2005speed, sawyer2016neutrino}. However, this artificial seeding is just a proxy for small perturbations induced by the vacuum Hamiltonian and an inhomogeneous matter distribution.

\begin{figure}
    \centering
    \includegraphics[width=\linewidth]{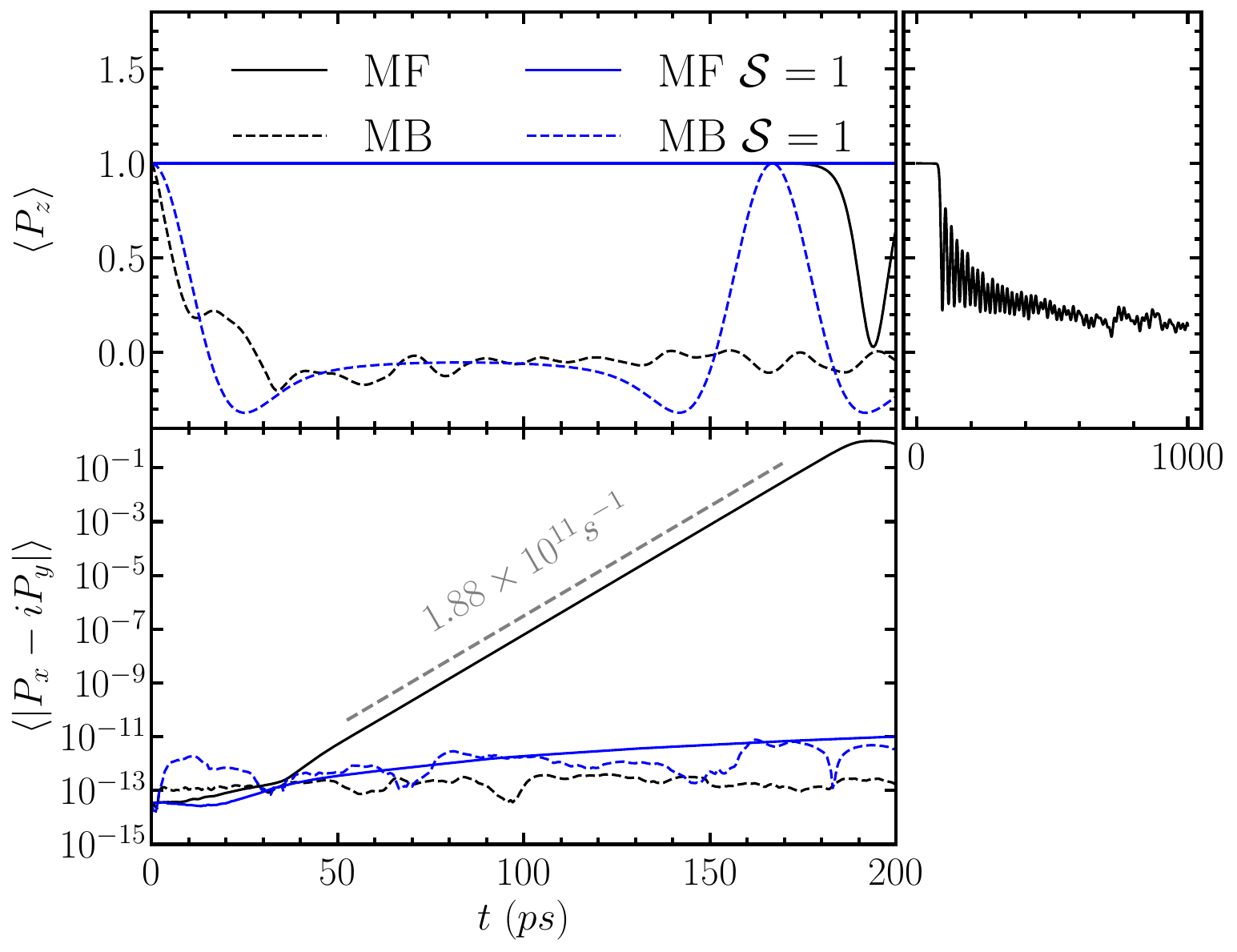}
    \caption{Inhomogeneous two-flavor, two-beam 20-particle system initially prepared with half muon and half electron neutrinos. \textbf{Top panel:} average of the density matrix component $\rho_{ee}$ over the first half of the sites. \textbf{Bottom panel:} quantum flavor coherence. Solid curves show MF results from setting BD=1 and dashed curves show MB results with full bond dimension. Black curves show results from using the inhomogeneous shape function and blue cuves show results from setting $\mathcal{S}=1$. \textbf{Top extended panel:} long-duration MF results using the {\tt Emu} code using $10^5$ particles on a periodic domain of size $10^3$ cm. Inhomogeneity allows rapidly settling to an equilibrium without constant or periodic behavior for both the MF and MB simulations.}
    
    %In the Inhomogenous mean-field (solid black lines) case, the instability grows exponentially, as muon neutrinos (red) rotate toward the pure electron state (blue) and a quantum superposition of electron and muon flavors (purple). The growth rate of the off-diagonal flavor component closely matches the prediction from linear stability theory. In contrast, the many-body inhomogeneous system (dashed black lines), prepared with the same initial flavor configuration and fully entangled at infinite bond dimension, exhibits suppression of FFI due to many-body entanglement, which effectively erases the inhomogeneous instability. Additionally, the behavior of this system under homogeneous conditions for both many-body and mean field cases (shown by dashed purple and dashed solid lines, respectively) show no growth of unstable mode. The \textbf{top extended panel} shows long-time mean-field inhomogeneous simulation, which asymptote toward a similar state as its corresponding many-body system, but on much longer timescales. These results demonstrate that inhomogeneity under many-body conditions triggers faster flavor equilibration compared to the mean-field and homoogenous case. \sr{[[]]}}

    \label{fig:Richers_Inhomo}
\end{figure}

%We now extend our investigation of the dynamics of the FFI in neutrino systems from MF to MB treatments. 
%\sr{\sout{The system is analyzed under both homogeneous and inhomogeneous configurations in the absence of a vacuum potential, to assess how spatial variations influence the onset and equilibration of FFIs, as shown in Fig.~\ref{fig:Richers_Inhomo}.}}

Our fiducial system consists of 20 particles. In the mean-field limit, the evolution of the flavor coherence exhibits a characteristic exponential growth indicative of a fast flavor instability (FFI). In Fig.~\ref{fig:Richers_Inhomo}, we demonstrate this behavior within our many-body framework at fixed BD=1 (solid black curve). Exponential growth emerges with a rate closely matching the analytic linear stability analysis (LSA) prediction, $\text{Im}(\omega) = 1.88\times10^{11}\,\text{s}^{-1}$ simulation~\cite{chakraborty2016self2, richers2021particle}. 
%= \frac{(m_2^2 - m_1^2)c^4}{2h\nu} + c k
Although the results shown here are based 20 sites, we demonstrate numerical convergence in the mean field limit for up to 1000 sites (see Appendix~\ref{app:convergence_symmetric}) that reproduce this analytic prediction within $3\%$ numerical error. 

The FFI saturates by $t \approx 200\,\text{ps}$, at which time $\langle \rho_{ee} \rangle$ transitions from 1 to 0.5 ($\langle P_z\rangle$ from 1 to 0), indicating flavor transformation to a mix of electron and muon neutrinos, before fluctuating back up again. These results mirror the behavior reported in Richers et al.~\cite{richers2021particle}. Although we converge on the growth rate well, we find that numerical errors grow in the CCNO results beyond a few hundred picoseconds, so we show the long-time depolarization of the neutrino distribution using an Emu simulation (in the top extended panel) using $2\times10^5$ particles and a domain size of $1000\,\mathrm{cm}$. At late times, the distribution approaches near complete mixing of neutrino flavors.

In contrast, the many-body inhomogeneous system (dashed black lines in Fig.~\ref{fig:Richers_Inhomo}) exhibits a rapid suppression of initial coherence growth (bottom panel). We use the full BD to ensure that all many-body effects are captured within the MPS representation and no external numerical truncation is involved to artificially ease computations. This allows our 20 particle system to serve as a minimal yet powerful benchmark for studying coherence, entanglement, and equilibration. The $x$ and $y$ components of the polarization vectors remain bounded at a level of $\sim 10^{-13}$ even at late times, while the $z$ components fluctuate near the equilibrated value of 0. This behavior reflects the emergence of decoherence due to many-body quantum entanglement, which effectively erases the collective instability and inhibits coherent flavor transformation. It can be clearly observed that the same equilibrated state is reached much earlier under MB dynamics than its MF counterpart. These findings align with previous work~\cite{kost2024once}, demonstrating that quantum decoherence overcomes mean-field coherence after long times, but at a rate that depends on the spacing between sites. In the many-body system, evolution saturates at a mixed state where coherent precession is lost and dynamics are dominated by decohering interactions among entangled particles.

\begin{figure}
    \centering
        \includegraphics[width=1.0\linewidth]{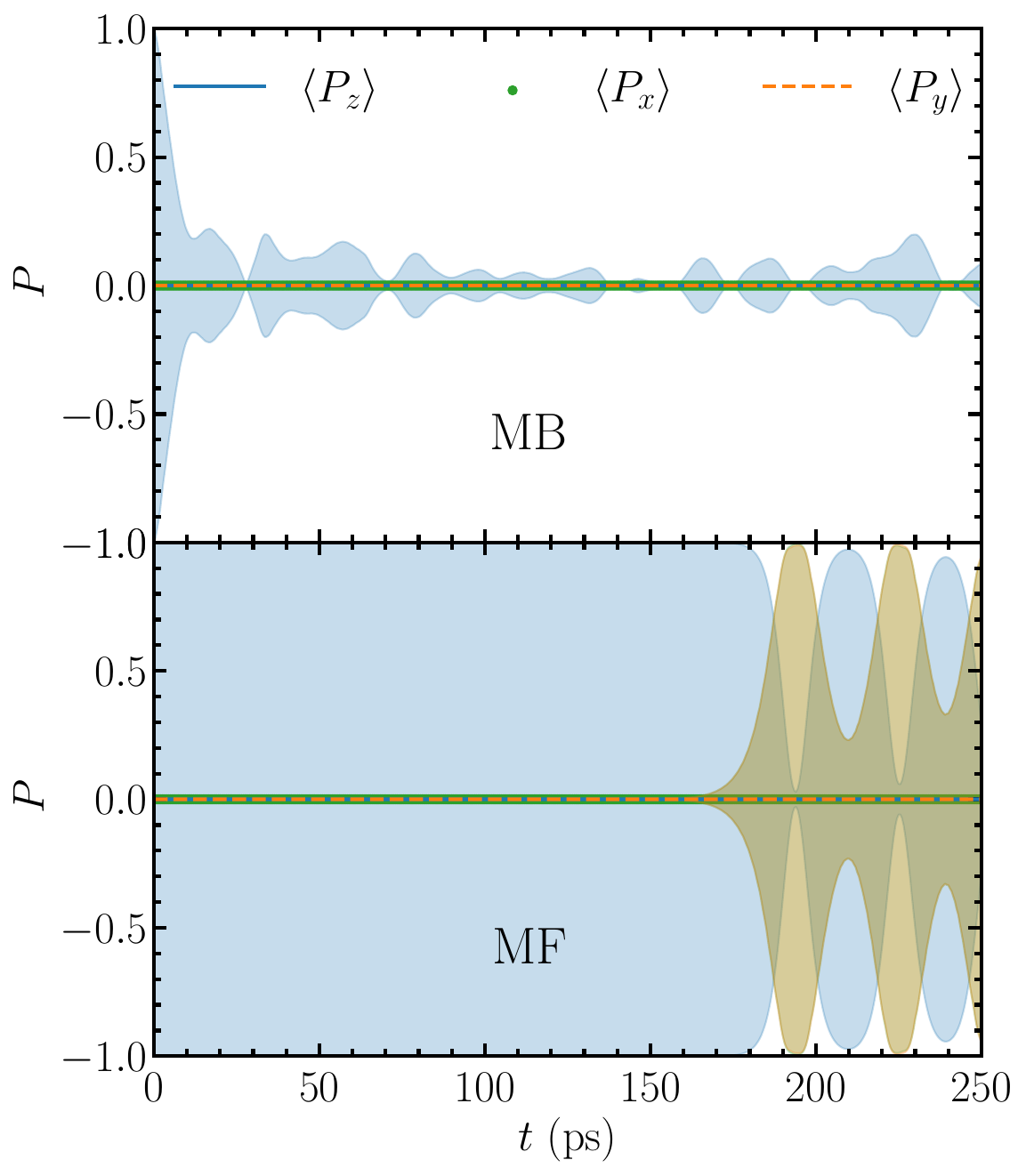}
    \caption{
Time evolution of the polarization vector components including their averages (solid constant lines) and the spread between their minimum and maximum values (shaded regions) for an all-sites 20-particle inhomogeneous system. \textbf{Upper panel:} Many-body simulation at infinite bond dimension. \textbf{Lower panel:} Mean-field simulation with bond dimension $=1$. In the many-body case, no exponential growth of transverse components is observed; instead, $\langle P_z \rangle$ undergoes sustained oscillations about zero, reflecting entanglement-driven decoherence. In contrast, the mean-field case shows each  beam polarized fully in $\pm1 \langle P_z \rangle$ showed with a fill region between the poles until a sharp rise in $P_x$ and $P_y$ at $t \sim 170$ ps, signaling the onset of flavor conversion through collective coherence. In both cases, the polarization averages of all components remain centered near zero, preserving total spin. }

    \label{fig:Sall_par_20}
\end{figure}
Figure~\ref{fig:Sall_par_20} displays the time evolution of the polarization vector components for all sites. The average values (opaque curves) remain at 0 for all components, as expected due to the flavor symmetry of the self-interaction Hamiltonian. The filled regions show the minimum and maximum values of the collection of particles.

% The average of each component for all sites \sr{\sout{(solid for $\langle P_z \rangle$, dashed for $\langle P_y \rangle$, and scatter for $\langle P_x \rangle$) exhibits behavior consistent with collective motion as seen by the oscillations after the MF FFI saturates}}\sr{[[what does this mean? What feature are you pointing out?]]}. While the fill bands between sitewise extrema highlight the oscillations spread across these extreme sites. \sr{[[This is a sentence fragment.]]}

In the many-body simulation (upper panel), the transverse components $\langle P_x \rangle$ and $\langle P_y \rangle$ remain near zero on average, with minimal sitewise spread, indicating a suppression of transverse polarization and the absence of exponential growth associated with FFI (as also seen in Fig.\ref{fig:Richers_Inhomo}). The $\langle P_z \rangle$ extrema oscillate about zero without further significant damping, consistent with the conclusion of \cite{martin2023equilibration} that the many-body evolution remains unitary and entangled, with equilibration occurring within a constrained subspace of the full Hilbert space. Overall, polarization averages are conserved, and the bounded late-time oscillations reflect entanglement-driven decoherence, in agreement with ~\cite{rrapaj2020exact}.

In contrast, the mean-field simulation (lower panel) exhibits exponential growth of $\langle P_x \rangle$ and $\langle P_y \rangle$, becoming visible after $\sim 170$ ps, signaling the saturation of the FFI. These dynamics reflect the buildup of flavor coherence, with larger transverse sitewise spread emerging after the instability. However, the initially fully polarized $\langle P_z \rangle$ continues oscillating between poles. We also note that these post-saturation oscillations in mean-field simulations are artifacts of the small particle number; simulations with more degrees of freedom allow neutrinos to decohere after a few coherent fluctuations, allowing faster depolarization, consistent with observations in Ref.~\cite{padilla2025flavor}.  

% \sr{[[I don't understand the point this paragraph is trying to make.]]} Comparison between many-body and mean-field results highlights the qualitative differences induced by entanglement. Unlike mean-field trajectories that evolve via smooth collective precession, the bounded late-time oscillations in many-body simulations reflect quantum fluctuations and entanglement-driven decoherence~\cite{rrapaj2020exact}. This pattern along with Fig. 5 of  Ref.~\cite{laraib2025many} hint towards the possibility that MB dynamics with persistent entanglement effects could dominate energetics even in larger systems.

\subsubsection{The Role of Inhomogeneity} \label{sec:inhomogeneity}
It is well understood that inhomogeneity allows modes of nonzero wavenumber to grow in the mean-field limit, but since the many-body depolarization is not a fast flavor instability, it is natural to wonder what role inhomogeneity has in many-body depolarization. We ran a modified version of the inhomogeneous Symmetric simulations, but setting $\mathcal{S}=1$ such that the neutrino self-interaction Hamiltonian is effectively homogeneous, as shown in Fig.~\ref{fig:Richers_Inhomo}. Note that this is a different configuration from that in Section~\ref{sec:Homogeneous FFI}. In the mean field limit, the two-beam model is not unstable to $k=0$ (i.e., homogeneous) modes, so there is no flavor transformation (blue solid curves). The MB equivalent (dashed blue curves) shows a rapid decrease in $\langle\rho_{ee}\rangle$ similar to the fiducial MB simulation. The large amount of symmetry in the Hamiltonian allows the system to oscillate periodically instead of fall into a chaotic equilibrium. This reinforces the idea that sufficient complexity in either spatial or angular structure is required to realize a quasi-static equilibrium.

\subsubsection{Flavor Asymmetry}\label{sec:Inhomo_MF}
\begin{figure}
    \centering
    \includegraphics[width=1.0\linewidth]{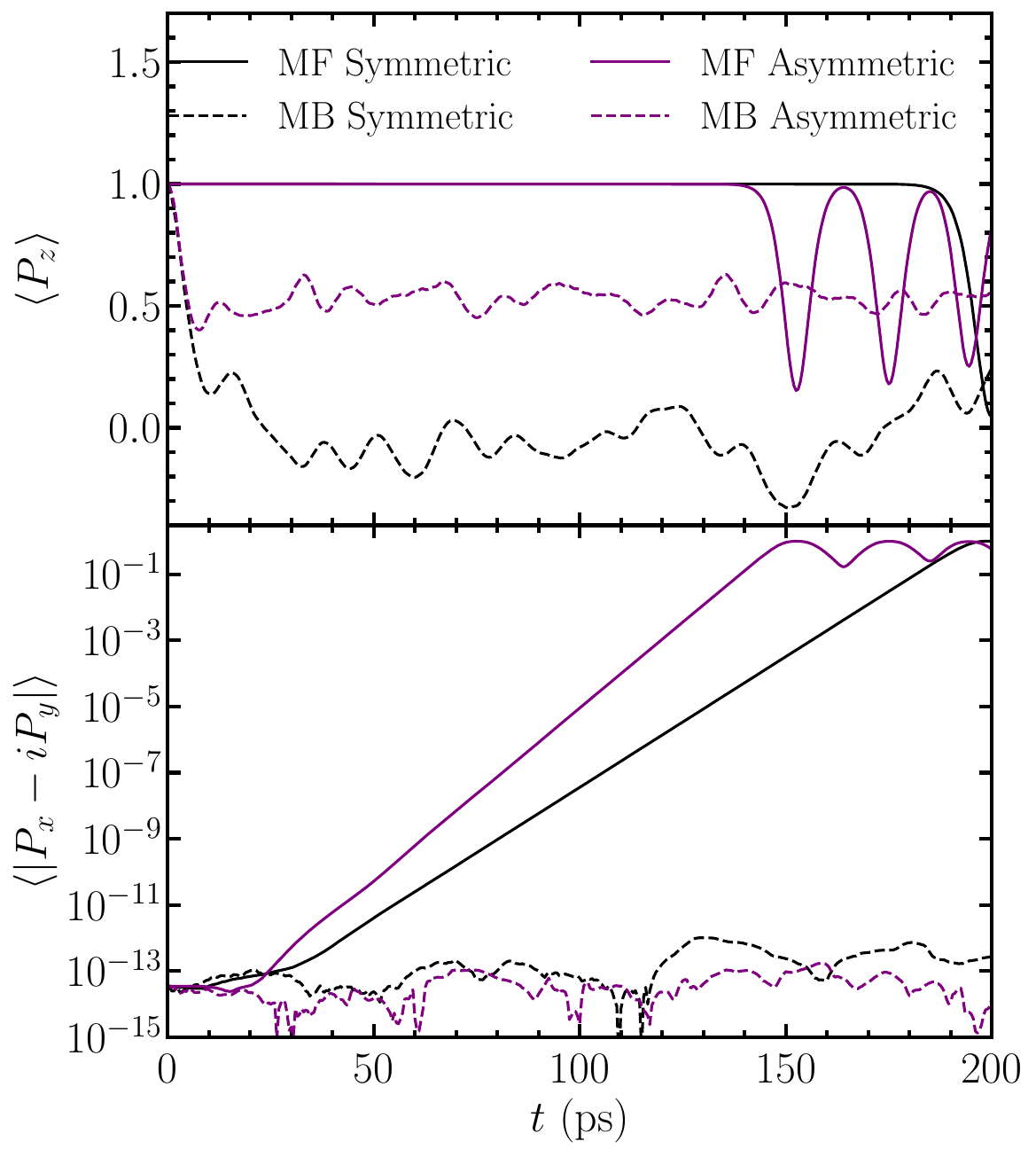}
    \caption{Evolution of the polarization vector components averaged over sites in the beam moving in the $+\hat{z}$ direction in terms of time scaled by $\mu=\sqrt{2}G_F n$. In the Symmetric simulations, there are seven sites in each neutrino beam. In Asymmetric simulations, there are seven sites in the $\nu_\mu$ beam and 14 sites in the $\nu_e$ beam. The many-body (MB) asymmetric system depolarizes faster than its symmetric counterpart and reaches the same asymptotic state as the mean-field (MF) simulations ($\langle P_z \rangle \approx 0.5$) significantly earlier. Fast flavor instability (FFI) also grows more rapidly under asymmetric conditions compared to the symmetric case.}
    \label{fig:sym_vs_asym}
\end{figure}

So far, all of our simulations have explored the special case of a Symmetric FFI when there are equal initial numbers of electron and muon neutrinos. We further explore the effect of flavor asymmetry in a two-beam model (Fig.~\ref{fig:sym_vs_asym}), where the number of electron neutrinos and similarly the number of electron neutrino sites is twice that of muon neutrinos (i.e., $N_{\rm sites}^{\nu_e} = 2\,N_{\rm sites}^{\nu_\mu}$). The domain for the Asymmetric simulations is the same as for the Symmetric simulations, as the wavelength of the fastest growing mode does not change.

In Fig.~\ref{fig:sym_vs_asym}, we show results from simulations using $N_{\rm sites}^{\nu_\mu} = 7$ for both the Symmetric and Asymmetric cases, and $N_{\rm sites}^{\nu_e} = 7$ for the Symmetric case, and $N_{\rm sites}^{\nu_e} = 14$ for the asymmetric case. In all cases particles are separated by $\Delta z=1/7\,\mathrm{cm}$ and have a shape function width of $w=1/7\,\mathrm{cm}$. We choose larger shape functions and spacing because the Asymmetric simulation would require 30 particles at our fiducial resolution, which we are not able to simulate with infinite bond dimension. This allows a direct comparison of many-body and mean-field dynamics for symmetric versus asymmetric flavor distributions under identical numerical choices. 

For both our Symmetric and Asymmetric configurations and in the mean-field limit, the flavor coherence exhibits a characteristic exponential growth indicative of a fast flavor instability (FFI). 
In Fig.~\ref{fig:sym_vs_asym}, we demonstrate this behavior within our many-body framework at fixed BD=1 for both the Symmetric and Asymmetric cases (solid curves). Exponential growth emerges with a rate consistent with the analytic linear stability analysis (LSA) predictions of $\text{Im}(\omega) = 1.88\times10^{11}\,\text{s}^{-1}$ for the Symmetric simulation and $2.66\times10^{11}\,\mathrm{s}^{-1}$ for the Asymmetric simulation~\cite{chakraborty2016self2, richers2021particle}. 
%= \frac{(m_2^2 - m_1^2)c^4}{2h\nu} + c k
Although the results shown here are based \sr{on} seven sites, we demonstrate numerical convergence in the mean field limit for up to 1000 sites (see Appendix~\ref{app:convergence_symmetric}) that reproduce this analytic prediction within $3\%$ numerical error. 

Upon saturation of the FFI, both MF simulations undergo rapid flavor fluctuations. As we discuss earlier in Section~\ref{sec:results_inhomo}, a high degree of symmetry or a small number of particles causes the post-saturation fluctuations to be very periodic, but a larger number of particles allows for the development of a chaotic system that settles to small fluctuations around an equilibrium. Although the first post-instability dip of the MF Symmetric simulation only dips down to $\langle P_z\rangle=0$, we know the long-term behavior of a larger MF system would follow the equilibration shown in the top right panel of Figure~\ref{fig:Richers_Inhomo}. Similarly, we see post-saturation fluctuations of the MF Asymmetric case around an equilibrium at a larger $\langle P_z\rangle$, consistent with the requirement to conserve lepton number.

The many-body Asymmetric system depolarizes initially at the same rate as the Symmetric simulation, but settles to an equilibrium sooner than the Symmetric simulation because less transformation is required to reach that equilibrium. However, many-body simulations--regardless of symmetry--develop flavor coherence without exhibiting the largest unstable mode, indicating the absence of FFI. Interestingly, while the equilibrium flavor coherence for individual particles is many orders of magnitudes smaller in the MB simulations, the flavor equilibrium is the same for the MF and MB simulations.

This suggests that, despite the disparate timescales, the net effect on the flavor distributions predicted by MF calculations could be similar to those predicted by the MB calculations. The difference is that MB calculations have a much larger Hilbert space, allowing the quantum state to explore paths to the final equilibrium that are much more efficient than the paths available to MF calculations, which by preventing many-body entanglement require large flavor coherence to achieve the same flavor distribution. However, this speculation will require more investigation to demonstrate its generality.

%==================================%
% Many Body: Finite Bond Dimension %
%==================================%
\subsubsection{Finite Bond Dimension}\label{sec:Finite_Bond_dimension}

\begin{figure}
    \centering
    \includegraphics[width=1.0\linewidth]{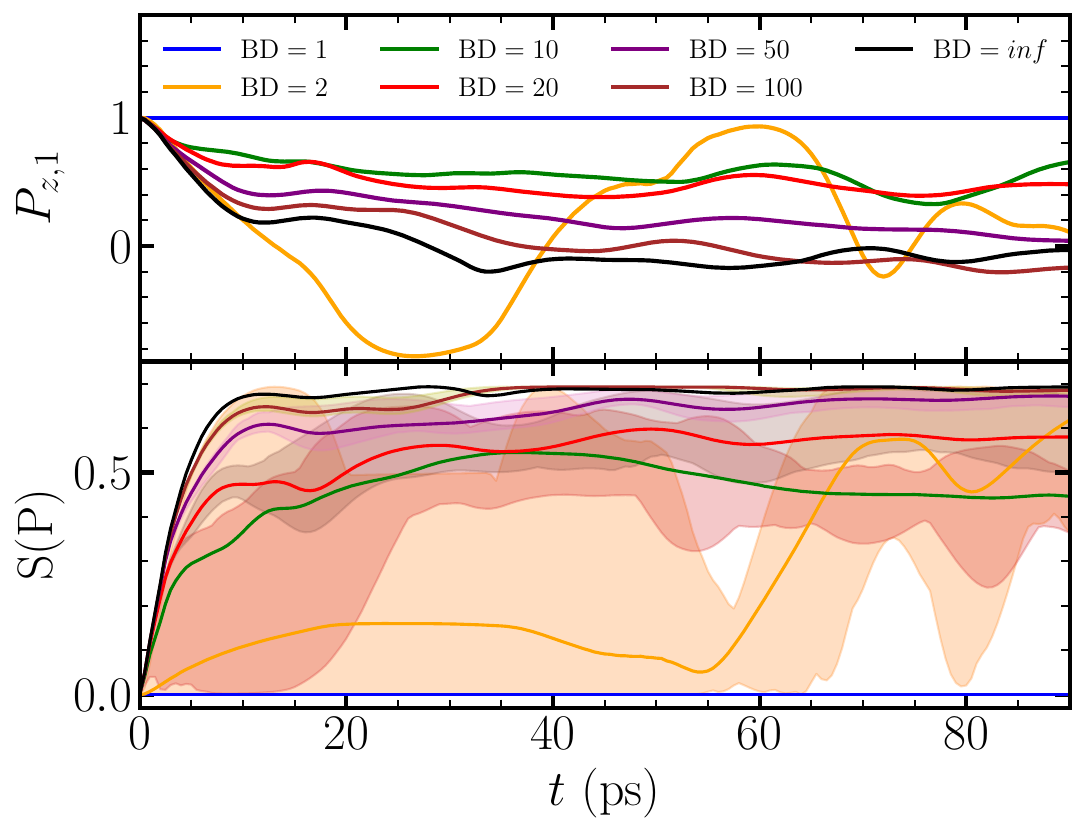}
    \caption{
    Time evolution for a 20-particle system under varying bond dimensions. \textbf{Upper panel:} $z$ component of the polarization vector at the first site.  \textbf{Lower panel:} Von Neumann entanglement entropy averaged over all sites (solid curves) and the spread between the minimum and maximum values (shaded areas). Entanglement entropy saturates with average values reaching $S\approx \ln 2$, close to the expected thermal limit for BD = $\infty$. At low BD, the entropy spread across sites is broader and the average entropy remains lower. A sufficiently large BD is essential for capturing full many-body coherence and flavor dynamics.}
    \label{fig:entropy_all_sites_all_par_periodic}
\end{figure}

The infinite bond dimension calculations contain numerical errors associated with the spatial discretization and the size of the timestep, but retained all entanglement information in the quantum state. We investigate the convergence of many-body neutrino evolution simulations with respect to \textit{bond dimension} (BD) in a 20-particle system, focusing on polarization, entanglement entropy, and flavor coherence to quantify the BD required to adequately capture many-body correlations. 

The top panel of Figure~\ref{fig:entropy_all_sites_all_par_periodic} shows the evolution of the first-site polarization $P^z_1$ under varying bond dimensions. At BD = 1, $P_z$ maintains a value of 1 since coherent oscillations develop much later, only after saturation of the FFI. For intermediate BDs, the dynamics gradually transition from mean-field-like behavior to the fully many-body regime. In particular, BD = 2 fails to capture the correct entanglement effects due to the strong truncation imposed by the tensor network. As the bond dimension increases (BD $\geq 10$), the time of the first polarization minimum $t_{\mathrm{min}}$ shifts slightly to later times and decreases in amplitude, followed by some irregular fluctuations, as observed in MF simulations after the FFI has saturated in late times.

The site-averaged von Neumann entropy in the infinite BD case, shown in the lower panel of Fig.~\ref{fig:entropy_all_sites_all_par_periodic}, quickly increases and saturates around $S \sim 0.67$, approaching the thermal limit ($\log 2 = 0.69$). At these higher bond dimensions, dynamics converge to the full many-body solution, with quantum entanglement also effectively damping the amplitude of local polarization $P_z$. As the BD decreases, the maximum entanglement entropy decreases, reflecting the erasure of entanglement by the SVD truncation.  In addition, lower BDs result in entropies that are highly non-uniform across sites, while at high BD the entropy becomes nearly uniform across sites. 

\begin{figure}
    \centering
    \includegraphics[width=1.0\linewidth]{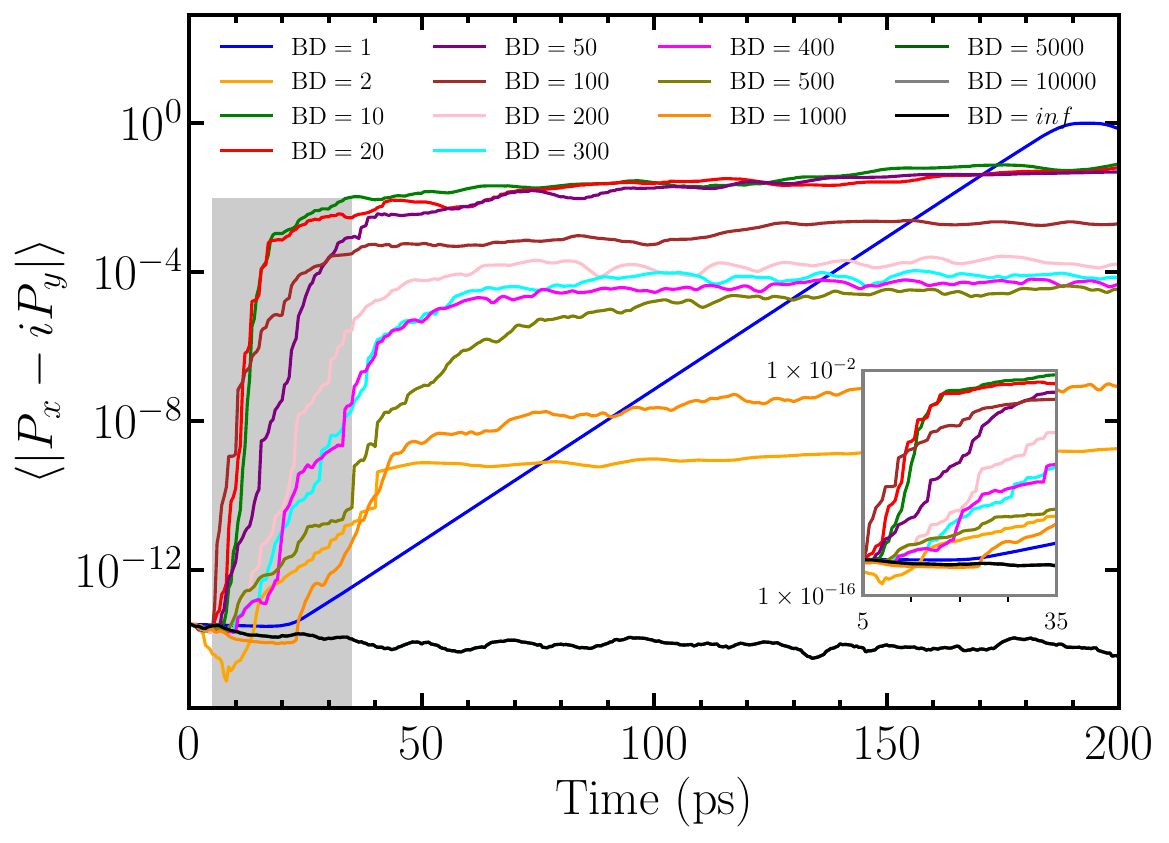}
    \caption{Time evolution of the average flavor coherence for a 20-particle system under periodic boundary conditions with varying bond dimensions and fixed timestep $\tau = 5 \times 10^{-13}$ s. The inset highlights the early-time behavior. For BD = 1, the extracted growth rate closely matches the linear stability prediction $\text{Im}(\omega) = 1.88 \times 10^{11}\,\mathrm{s}^{-1}$ with a numerical error of 0.17\%. For BD~$\geq 10$, the observed growth rates diverge significantly, reaching up to $2.5 \times 10^{12}\,\mathrm{s}^{-1}$ (BD=10). BD=100000 overlaps on top of BDinf showing full convergence with MB results and no growth of the unstable mode.}
    \label{fig:rho_emu_20_par}
\end{figure}

The dynamics of the average flavor coherence shown in Figure~\ref{fig:rho_emu_20_par}, exhibit a strong dependence on bond dimension. In the mean-field limit (BD = 1), flavor coherence grows exponentially and in the infinite bond dimension case values remain small, as demonstrated in  Fig.~\ref{fig:Richers_Inhomo}. However, we also observed that for intermediate bond dimensions, it evolves in a qualitatively different manner, much faster than the FFI predicted by mean-field theory. As BD increases from 1, the growth rate increases dramatically, reaching up to $\sim 2.50 \times 10^{12}\, \mathrm{s}^{-1}$ for BD = 10, more than an order of magnitude larger than the linear stability prediction. However, as the BD increases past 5,000, the flavor coherence growth is suppressed. At these high BDs (beyond BD= 10,000) where many-body effects have fully kicked in, further increasing BD for MB simulations shows a smaller asymptotic values of flavor coherence, than its MF counterpart. It is worth emphasizing that the errors in $\rho_{e\mu}$ \textit{grow} with increasing bond dimension until there is a sufficiently large bond dimension to prevent growth of $\rho_{e\mu}$ at all. Thus, bond dimension truncations need to be done with great care.
  
% \sr{\sout{We also observed that for intermediate bond dimensions, $\rho_{e\mu}$ evolves in a qualitatively different manner, much faster than the FFI predicted by mean-field theory; nonetheless, the diagonal polarization $P_z$ and entanglement entropy $S$ do not differ significantly between BD = 100 and the infinite-BD case. [[This was already said, I think]]}}

These findings emphasize the necessity of sufficiently large bond dimensions to accurately simulate quantum decoherence, flavor equilibration, and thermal-like entanglement saturation in dense neutrino systems. High-BD simulations, however, scale exponentially with system size, meaning that larger systems require proportionally higher BD to fully capture correlations and entanglement. This consideration motivates the choice of $N=20$ particles, providing a practical balance between computational cost and fidelity while still accurately representing genuine many-body dynamics.

% \begin{table}[h]
% \centering
% \caption{Extracted growth rates $\mathrm{Im}(\omega)$ for different bond dimensions (BD) for 20 and 100 particle systems. The analytical mean-field prediction is $\text{Im}(\omega) = 1.88 \times 10^{11}$ s$^{-1}$. Errors are computed relative to this value.}
% \label{tab:rho_emu_growth_rates}
% \begin{tabular}{|c|cc|c|c|}
% \hline
% \textbf{Particles} & \textbf{BD} & \textbf{Time Interval (ps)} & \textbf{$\mathrm{Im}(\omega)$ (s$^{-1}$)} & \textbf{Error (\%)} \\
% \hline
% \multirow{20} 
%   & 1   & 48.00 - 163.50 & $1.88 \times 10^{11}$ & 0.19  \\
%   & 2   & 47.00 - 44.50 & $1.72 \times 10^{11}$ & 8.67  \\
%   & 10  & 2.00 - 8.00   & $2.10 \times 10^{12}$ & 1014.61 \\
%   & 20  & 3.00 - 13.00  & $1.75 \times 10^{12}$ & 830.57 \\
%   & 50  & 1.00 - 11.00  & $1.72 \times 10^{12}$ & 811.31 \\
%   & 100 & 2.00 - 9.00   & $1.02 \times 10^{12}$ & 444.11 \\
% \hline
% \multirow{100} 
%   & 1   & 72.00 - 184.00 & $1.93 \times 10^{11}$ & 2.61 \\
%   & 2   & 28.00 - 38.50  & $1.84 \times 10^{11}$ & 2.50 \\
%   & 10  & 1.00 - 5.00    & $1.04 \times 10^{12}$ & 449.94 \\
%   & 50  & 2.50 - 14.00   & $1.55 \times 10^{12}$ & 723.68 \\
%   & 100 & 1.00 - 14.00   & $8.70 \times 10^{11}$ & 361.89 \\
% \hline
% \end{tabular}
% \label{table:rho_emu errors}
% \end{table}

\subsubsection{Open vs Closed systems}
\label{sec:open_closed}
Existing many-body (MB) studies of collective neutrino oscillations are typically constrained by closed (periodic) boundaries and idealized symmetry assumptions. To probe how these simplifications influence the dynamics, we systematically relax each of these assumptions, beginning with the role of boundary conditions.  Our first question is simple but fundamental: Do open and closed boundary conditions produce qualitatively different many-body behavior? 

\begin{figure}
    \centering
    \includegraphics[width=1.0\linewidth]{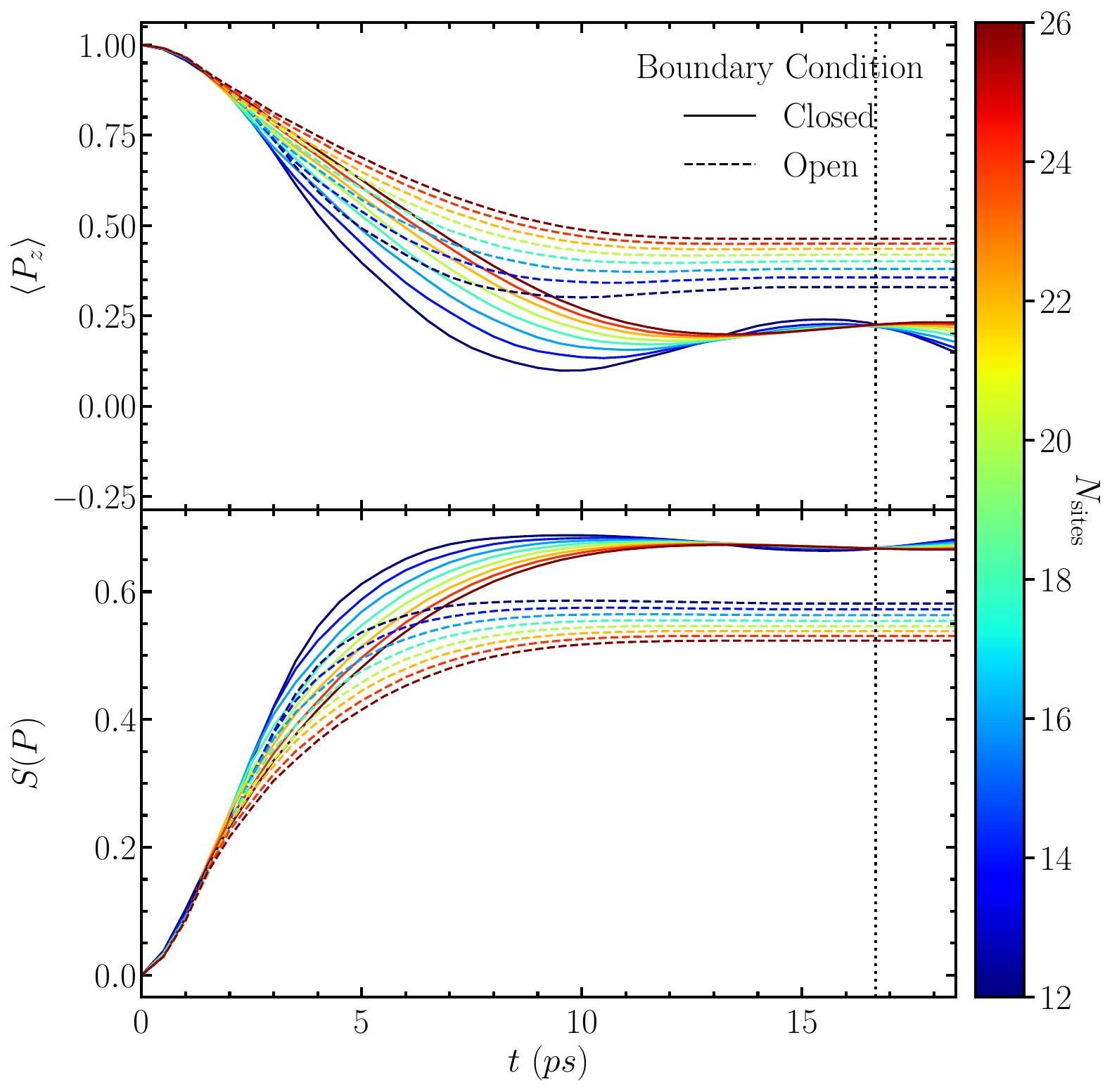}
    \caption{Time evolution of a flavor-symmetric inhomogeneous neutrino system with superimposed initial conditions, comparing open (dashed) and closed (solid) boundary conditions for varying $N_{\rm sites}$. In Open BC simulations, flavor transformation ceases because particles interact only once and do not re-encounter each other, leading to lower final entanglement entropy. In Closed BC simulations, repeated interactions from periodic wrapping result in higher final entanglement and more sustained flavor transformation. Dashed black line shows one lap time.}
    \label{fig:Superimposed_flavor_symmetric}
\end{figure}
Figure~\ref{fig:Superimposed_flavor_symmetric} compares the flavor evolution of an initially flavor-symmetric neutrino system under open (dashed) and closed (solid) boundaries. In both cases, we place the initial locations of the electron neutrinos ($\nu_e$) and muon neutrinos ($\nu_\mu$) at the same locations evenly spaced between $z=0$ and $z=1\,\mathrm{cm}$ (the $N_\mathrm{sites}=20$ Closed simulation is identical to the Fiducial simulation). However, in the the Open simulations, neutrinos continue moving toward $z=\pm\infty$ without wrapping around the periodic boundaries. This means that after one lap time $t=16.678$ ps (given by vertical black dashed line) the trains of sites are spatially separated and no longer interact, so the Hamiltonian is zero. This comparison is especially motivated by the streaming behavior of neutrinos outside of the neutrinosphere, where any pair of neutrinos interacts only once, in contrast to the debated realism of the repeated interactions resulting from Closed BCs \cite{shalgar2023we}.

Both boundary conditions display qualitatively similar initial behavior, since the sites are identically located and interactions are very similar.  However, as evolution proceeds, their behaviors diverge. In the open system, neutrinos near the edges escape and stop interacting, causing the entropy growth and flavor transformation to cease within one lap time. This leads to a much faster freeze out to $\langle P_z\rangle\!\approx\!0.5$ for larger $N_\mathrm{sites}$. In contrast, in the closed (periodic) system, particles repeatedly interact through boundary wrapping, maintaining collective correlations longer and sustaining stronger many-body entanglement. As discussed in Ref.~\cite{laraib2025many}, a larger $N_\mathrm{sites}$ leads to slower flavor transformation and entropy growth. This trend is reflected in both the Open and Periodic boundary condition simulations.
% \sr{[[Do we even want to discuss tmin? The open BC simulations don't even really have a tmin.]] \sout{The apparent independence of $t_{\rm min}$ on system size in the open case indicates that once the beams have streamed past each other, the system effectively decouples. This contrasts with the closed case, where $t_{\rm min}$ shifts to later times with increasing $N_{\rm sites}$ due to recurring encounters. Importantly, this comparison highlights that boundary conditions alone can dramatically alter the persistence and strength of many-body effects, even when initial states are identical.}}

\subsubsection{
Superimposed vs. Separated Initial Conditions} \label{sec:results_IC_spatial_config}

The above simulations of the MB system with open boundary conditions assumed that neutrinos begin superimposed with no flavor coherence or many-body entanglement. However, it is not clear that this assumption is a good representation of the behavior of neutrinos in astrophysical environments, since neutrinos may propagate before undergoing flavor transformation.

%Figure~\ref{fig:Superimposed_flavor_symmetric} compares the flavor evolution of an initially flavor-symmetric neutrino system under open (dashed) and closed (solid) boundaries. In both cases, \sr{we place the initial locations of the \sout{the initial state places}} electron neutrinos ($\nu_e$) and muon neutrinos ($\nu_\mu$) on opposite halves of the domain, \sr{such that $x\in(-1,0)$ contains ten evenly-spaced right-moving electron neutrino sites, and $x\in(0,1)$ contains ten left-moving muon neutrino sites. Their separation means they do not interact initially, but as the trains of sites move past each other they \sout{allowing them to}} interact as counter-streaming beams. This comparison is especially motivated to mimic the streaming behavior of neutrinos \sr{outside of \sout{near}} the neutrinosphere, where \sr{any pair of neutrinos interacts only once \sout{Open BCs suppress artificial self-interactions and ensure causal, physically consistent flavor evolution}}, in contrast to the \sr{\sout{widely debated} debatable realism of the repeated interactions resulting from} Closed BCs \cite{shalgar2023we}.

\begin{figure}
\centering
\includegraphics[width=1.0\linewidth]{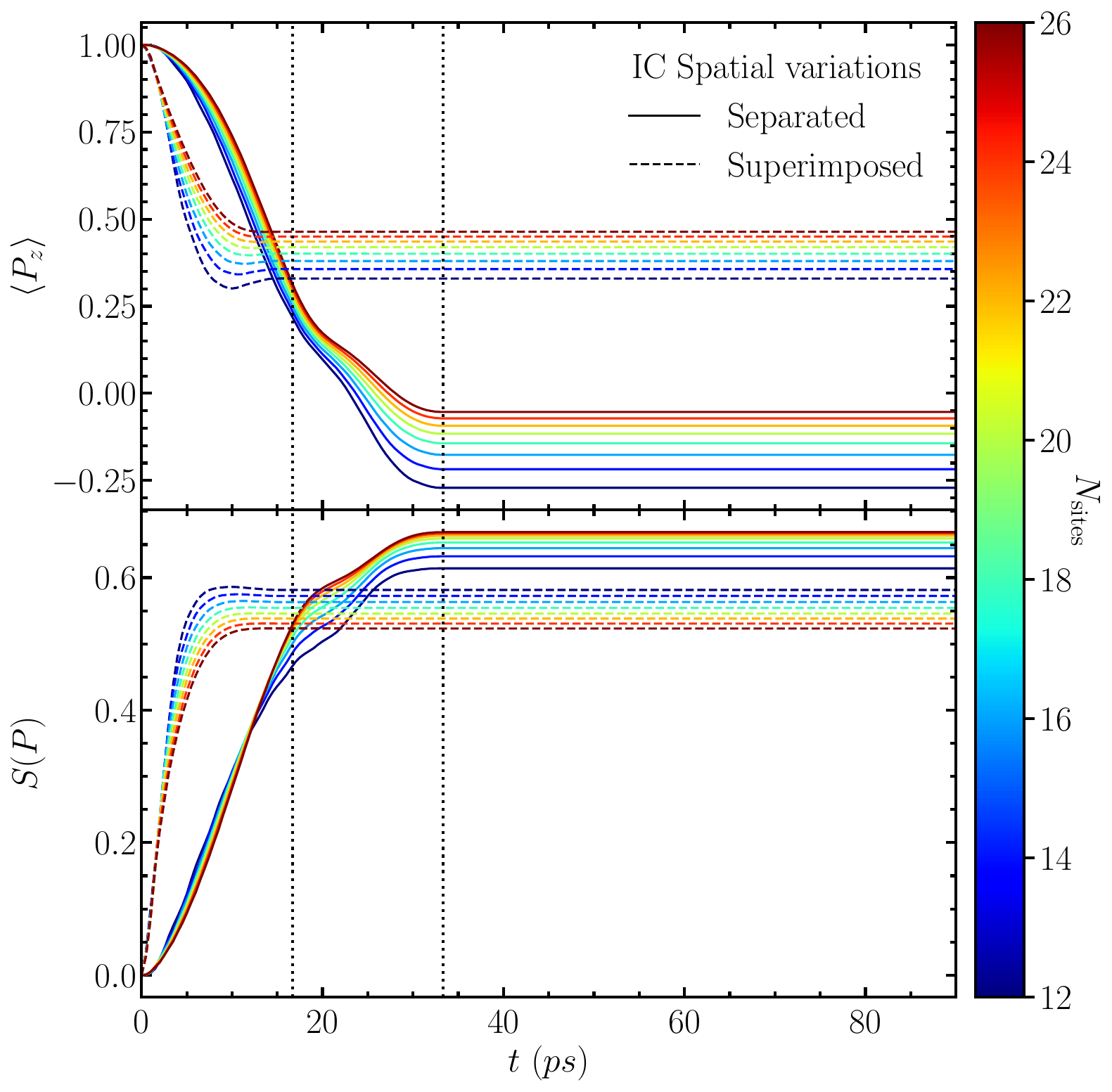}
\caption{Time evolution of a flavor-symmetric inhomogeneous neutrino system with open boundary conditions, comparing Separated (solid) and Superimposed (dashed) initial conditions. We show the average $z$ component of the single-site polarization vectors (top panel) and entanglement entropy (bottom panel) under open boundary conditions for different initial spatial configurations. In the Superimposed configuration (dashed), where $\nu_e$ and $\nu_\mu$ occupy opposite halves of the domain, many-body effects weaken as $N_{\rm sites}$ increases. In the Separated configuration (solid), where the muon neutrino beam is initially positioned outside the domain and moves past the electron neutrinos, many-body effects are comparable to the superimposed case within the first lap time but strengthen with increasing $N_{\rm sites}$ after one lap. Vertical lines mark the approximate lap times: the Superimposed case saturates once interactions vanish after the first lap, whereas the Separated case continues interacting beyond the first lap, with the system fully interacted by the second.}
\label{fig:OpenBC_flavor_sym_spatial_config}
\end{figure}
To test an alternate toy model in which neutrinos do not immediately change flavor upon creation (i.e., the start of the simulation), we designed two sets of experiments under Open BCs. Figure~\ref{fig:OpenBC_flavor_sym_spatial_config} compares these two setups:  
(1) the superimposed configuration (dashed), where $\nu_e$ and $\nu_\mu$ are initially both distributed in a 1 cm domain (as in Section~\ref{sec:open_closed}), and  
(2) the separated configuration (solid), where the $\nu_e$ beam occupies the region $z\in[0,1]\,\mathrm{cm}$, while an equally numbered $\nu_\mu$ beam is initially positioned  in the region $z\in[1,2]\,\mathrm{cm}$ and allowed to move past the other beam, briefly interacting with the electron neutrinos as they pass. Both begin in flavor-symmetric initial states, but their interaction durations differ dramatically.

At first, the results seem intuitive: in the superimposed case, where beams interact strongly at $t=0$, many-body correlations grow rapidly, producing a steeper early-time slope than in the separated case. This slope decreases as $N_{\rm sites}$ increases, consistent with the trend identified in our previous work\cite{laraib2025many} for closed BCs, indicating that this effect is not an artifact of single interactions.
% The $t_{\rm min}$ for flavor transformation remains roughly constant with $N_{\rm sites}$, consistent with the open-boundary trend seen in Figure~\ref{fig:Superimposed_flavor_symmetric}. \sr{[[Should we even talk about a tmin? I don't see a minimum on all of the curves.]]} 

The separated configuration, however, tells a more surprising story.  Although both beams always remain inside the computational domain, the slower initial evolution of the Separated curve is the result of partial overlap of beams. During this stage, the evolution resembles the superimposed case in the sense that the same trend—larger $N_{\rm sites}$ leading to slower growth—still holds, though with a reduced overall rate due to the smaller overlap region. In the superimposed setup, this early growth is followed by an eventual stagnation of entanglement as the beams separate. In contrast, after one full lap ($16.678$ ps), the separated configuration exhibits the opposite behavior: the entanglement strengthens with increasing $N_{\rm sites}$ rather than diminishing. A plateau followed by a second dip in polarization (smaller dip with increased $N_{\rm sites}$) and higher late-time entropy (also with increasing $N_{\rm sites}$) appear, revealing that the initial separation introduces new, long-lived collective behavior not seen in the superimposed setup. Due to these extended interaction times, entropy saturates to a higher maximal value as compared to superimposed BCs.

This suggests that even when open systems initially behave like their periodic counterparts, the timing and extent of the initial overlap region—specifically, whether the interaction begins in a fully superimposed or only partially overlapped state—can qualitatively modify the long-time behavior. However, both of these setups interact for a relatively short time, and there is only about a single order of magnitude separation between the size of a site and the size of the full train of neutrinos.

\section{Conclusions}\label{sec:conclusion}

The inhomogeneous tensor-network framework presented in this work allows us to consistently treat the effects of periodicity, spatial configuration, flavor asymmetry, system size, and geometry on many-body (MB) neutrino flavor evolution in the context of the neutrino fast flavor instability (FFI). These simulations together illuminate how boundary conditions, geometry, and interaction duration collectively determine the emergence and longevity of many-body effects in dense neutrino systems. 

Our studies show that MB effects fundamentally modify FFI dynamics by replacing exponential growth with rapid depolarization, bounding single-site flavor coherence under both homogeneous and inhomogeneous conditions to very small values. Inhomogeneous systems equilibrate rapidly through entanglement-induced decoherence, while homogeneous ones exhibit periodic behavior. Flavor asymmetry (i.e., initial states with more electron neutrinos than muon neutrinos) accelerates depolarization just as it accelerates the growth of the FFI in the mean-field limit, although MB effects are still sufficiently rapid to cut off the FFI in our tests.

Through adjusting the bond dimension we bridge the MF and MB regimes and highlight that sufficient capacity to capture entanglement in the quantum state is essential to accurately model the interplay between coherence growth, decoherence, and flavor equilibration. At low bond dimension ($\mathrm{BD}=1$), the system reproduces mean-field effects, including the FFI and flavor fluctuations after saturation of the FFI. In the infinite bond dimension limit, the system fully resolves MB correlations, showing entanglement-driven suppression of FFIs, bounded late-time fluctuations, and equilibration. At intermediate bond dimensions, evolution of the flavor content and entanglement entropy of the beams is between the mean-field and MB limits. However, single-site flavor coherence quickly shoots up to large values, even though it remains initially small in both the mean-field and MB limits. Rapid single-site flavor coherence growth and a wide spread of entanglement entropies between different sites can be used as a metrics of insufficient bond dimension.

We systematically relax imposed assumptions of boundaries and symmetries in the system and evaluate whether these assumptions change the underlying dynamics fundamentally. Open boundary conditions allow entanglement and flavor evolution to develop naturally without the risk of pairs of neutrinos circling around a periodic domain and interacting multiple times. Under these conditions, initially superimposed configurations show strong entanglement and fast flavor equilibration, similar to the results of periodic simulations, until the trains of neutrinos separate and stop interacting. These results suggest that periodic-like behavior could be recovered from systems with open boundary conditions of the beams are sufficiently long. However, initially separated configurations sustain interactions that are slow to ramp up as the opposing beams come into contact with each other, but also result in longer interaction times. Decreasing the spatial separation between sites in initially separated setups decreases the rate at which flavor changes, but increases the rate at which entanglement builds up, contrary to the Superimposed systems. It is not clear which initial conditions, if any, accurately represent the MB dynamics that would occur in astrophysical environments.
%This is opposite to the trend observed when varying the domain length under fixed site spacing. 
% \sr{\sout{Our results demonstrate that open systems can reproduce closed-system MB behavior when the interaction region is sufficiently large, but only for configurations with sustained beam overlap. However, all MB effects emerge significantly earlier than in mean-field approximations, underscoring the need to control spatial configuration, system size, and boundary conditions to model dense neutrino systems accurately.}}

% Together, these results suggest that open systems can indeed recover closed-system MB behavior if the neutrinos begin superimposed in flavor eigenstates and the spatial extent of the distribution of neutrinos is sufficiently large. However, systems that begin spatially isolated and must advect to interact, behaving in a qualitatively different manner, such that the final amount of flavor transformation increases with $N_\mathrm{sites}$ (i.e., simulation fidelity). This raises the question as to which setup is more representative of neutrino flavor transformation in the highly inhomogeneous and anisotropic environments of core-collapse supernovae and neutron star mergers.  

While this study establishes a robust foundation for scalable, high-fidelity MB simulations of neutrino flavor evolution, several limitations remain that outline natural avenues for future work. The present framework is restricted to a two-beam, two-flavor system and neglects non-forward scattering terms for general momentum-transfer between neutrino pairs \cite{cirigliano2024neutrino}, which would all be necessary to capture the full complexity of flavor transport in realistic astrophysical environments. Additionally, extending this approach to higher-dimensional, multi-beam, multi-angle configurations would allow the inclusion of richer angular structures and many-body entanglement patterns relevant to global supernova and merger geometries. Complementary investigations have introduced the wave-packet formalism as a separate improvement to the description of neutrino interactions \cite{cervia2025interactions}. Its extension to larger many-body systems and evaluation within an inhomogeneous setup remains an open challenge. Beyond these physical extensions, the tensor-network framework developed here offers a promising platform for benchmarking emerging quantum simulators and hybrid classical–quantum algorithms designed to capture entanglement-driven neutrino dynamics.

\section{Acknowledgements}
This work benefited from a number of insightful conversations with Baha Balantekin, Michael Cervia, and Ermal Rrapaj. ZL and SR gratefully acknowledge support by the National Science Foundation under Award No. PHY-2412683.

\bibliography{apssamp}
% Produces the bibliography via BibTeX.
\appendix
\section{Mean field tests}

\subsection{Vacuum Oscillations}\label{sec:classical_vac_osc}
We evolve a collection of six neutrinos, half of which start in the $| \nu_\mu \rangle$ state and the other half of which start in the $|\nu_e\rangle$ state, i.e., \( |\Psi_0\rangle = \left(\bigotimes_{n=1}^{(N_\mathrm{sites}/2)} |\downarrow \rangle\right) \otimes \left(\bigotimes_{m=1}^{(N_\mathrm{sites}/2)} |\uparrow \rangle\right) \). We set $N_i=0$ for each site $i$ such that the self-interaction Hamiltonian is zero. We set $|\vec{p}_i|=1\,\mathrm{MeV}$, $m_1=0\,\mathrm{eV}$, $m_2=8.60\times10^{-3}\,\mathrm{eV}$ and $\theta=\pi/4$ such that \( \vec{B} = [1, 0, 0] \) and all neutrinos oscillate with a frequency of \( \omega = \Delta m^2/2E = 5.61\times10^{4}\,\mathrm{s}^{-1}\) in the plane spanned by the y- and z-axes. We evolve for a total time of $1.94\times10^{-4}\,\mathrm{s}$ using 200 time steps.

\begin{figure}
    \centering
        \includegraphics[width=\linewidth]{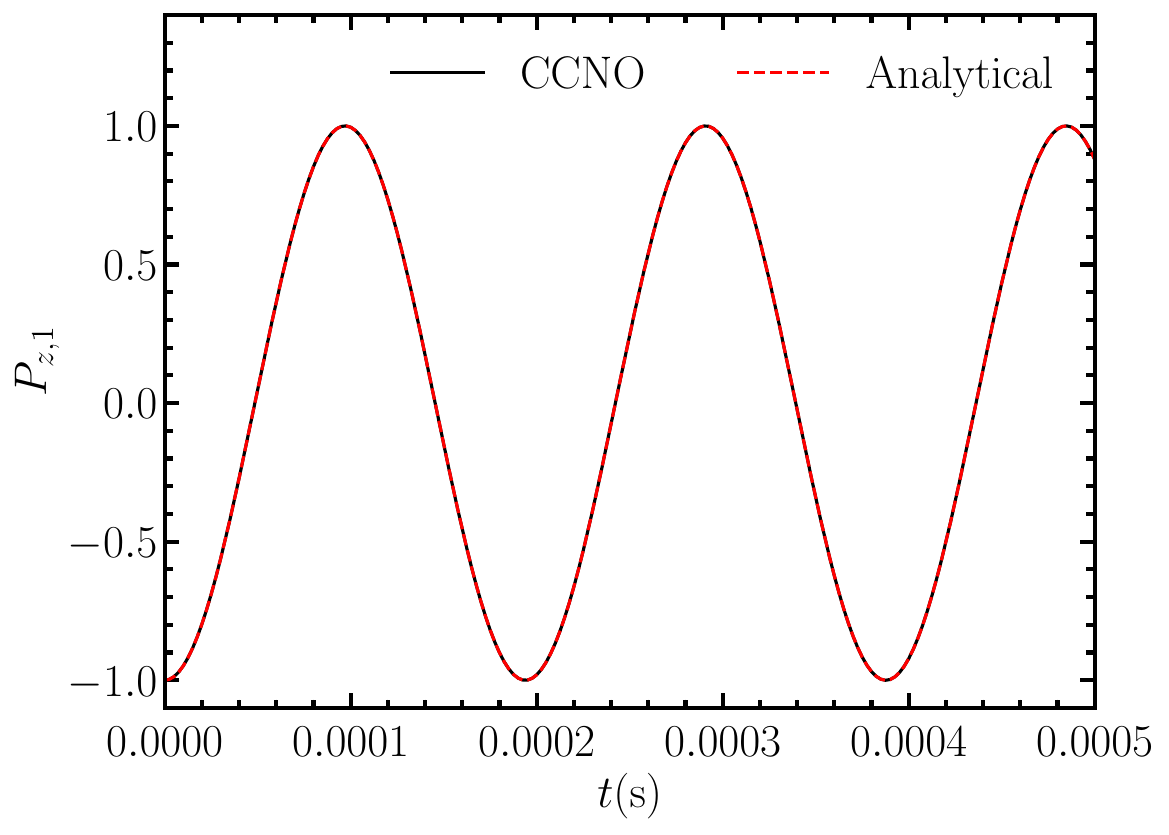}
    \caption{Vacuum oscillation test. Plotted here is the evolution of a single neutrino initially in the $|\nu_\mu\rangle$ state. Results match analytic predictions to floating point precision with a maximum error of $4.2\times10^{-15}$.}
    \label{fig:vac_osc}
\end{figure}

Under these conditions, $P_{z,i} (t) = -\cos(\omega_i t)$ for the sites starting in the muon state (and negative of this for sites starting in the electron state). The theoretical and simulated results for one of these particles is shown in Fig.~(\ref{fig:vac_osc}).

%%%%%%%%%%%
% Bipolar %
%%%%%%%%%%%
\subsection{Bipolar Oscillations}
\begin{figure}
    \centering
\includegraphics[width=\linewidth]{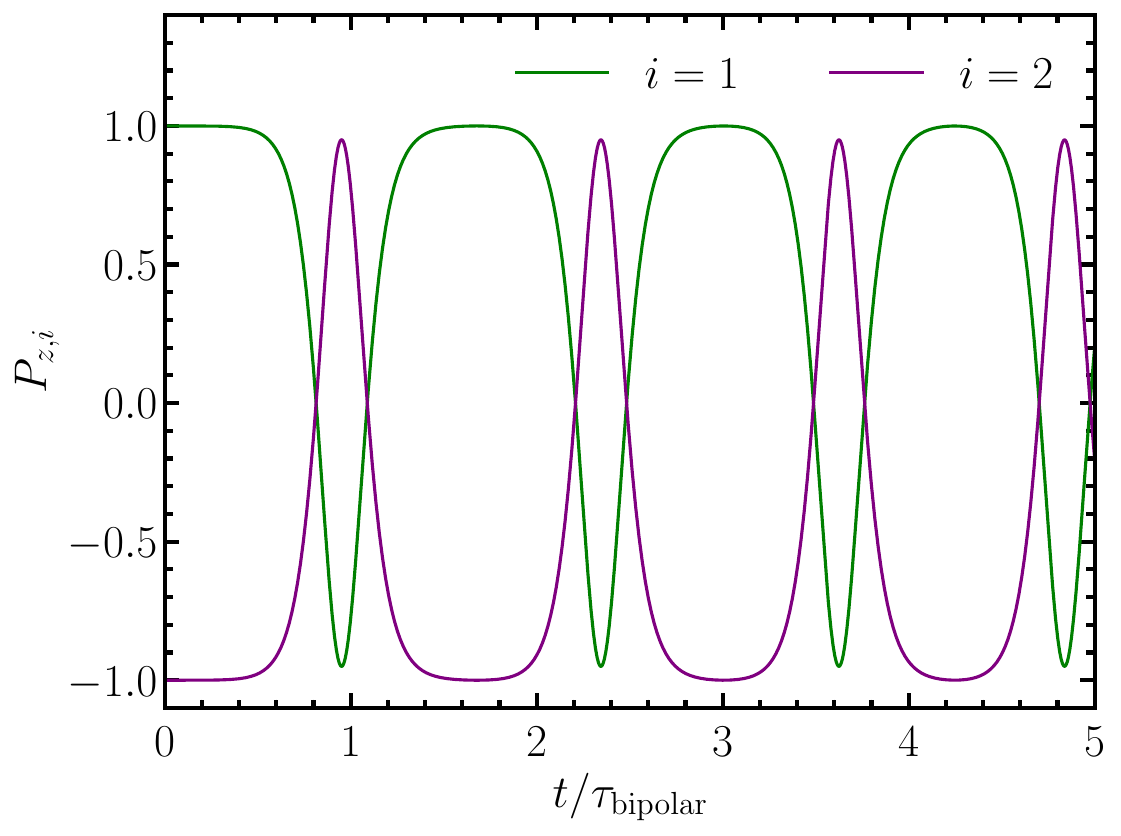}
    \caption{An isotropic one-zone bipolar oscillation test using two counter-propagating beams. We set the mixing angle $\theta_{12}$ such that $H_{\text{SI}}/H_{\text{vac}}$ = 10 with characteristic oscillation time \( \tau_\mathrm{bipolar} = 8.96 \times 10^{-4} \) s. This plot can be compared with Figure 13 in \cite{richers2021particle}.}
    \label{fig:Richers_bipolar_MF}
\end{figure}

We reproduce homogeneous and isotropic bipolar oscillations introduced in \cite{hannestad2006self} with parameters matching those of \cite{richers2021particle}. We use neutrino masses of $m_1 = 8.50\times10^{-3}\,\mathrm{eV}$ and $m_2 = 0\,\mathrm{eV}$, corresponding to the inverted mass ordering, and set the mixing angle to $\theta_{12} = 0.01$. We choose particle weights to ensure the electron and muon neutrino number densities are \( n_{\nu_e} = n_{\nu_\mu} = 10(m_2 - m_1)^2c^4/(2 \sqrt{2} G_F E) \) with energy \( E = 50 \) MeV. These choices correspond to the $\delta \omega  / \mu = 0.1$ from \cite{roggero2021entanglement} with a characteristic oscillation time of \( \tau_\mathrm{bipolar} = 8.96 \times 10^{-4}\,\mathrm{s} \) (see Equation 21 in \cite{hannestad2006self}). We set the initial state to $| \nu_e \nu_\mu\rangle$. The results are plotted in Fig.~\ref{fig:Richers_bipolar_MF} and show clear bipolar oscillations on the bipolar timescale. This is a challenging test for the time integrator, is it must correctly evolve small perturbations over many timesteps during the plateu phases of the bipolar oscillations. We use a simple forward-Euler time integrator, so we need to use quite small timesteps to prevent the time between successive oscillations from decreasing. Specifically, we evolve for a total time of $10^{-2}\,\mathrm{s}$ using timesteps of $10^{-8}\,\mathrm{s}$.

%=======================%
% Symmetric Convergence %
%=======================%
\subsection{Symmetric Fast Flavor Instability Convergence}
\label{app:convergence_symmetric}
\begin{figure}
    \centering        
    \includegraphics[width=1.0\linewidth]{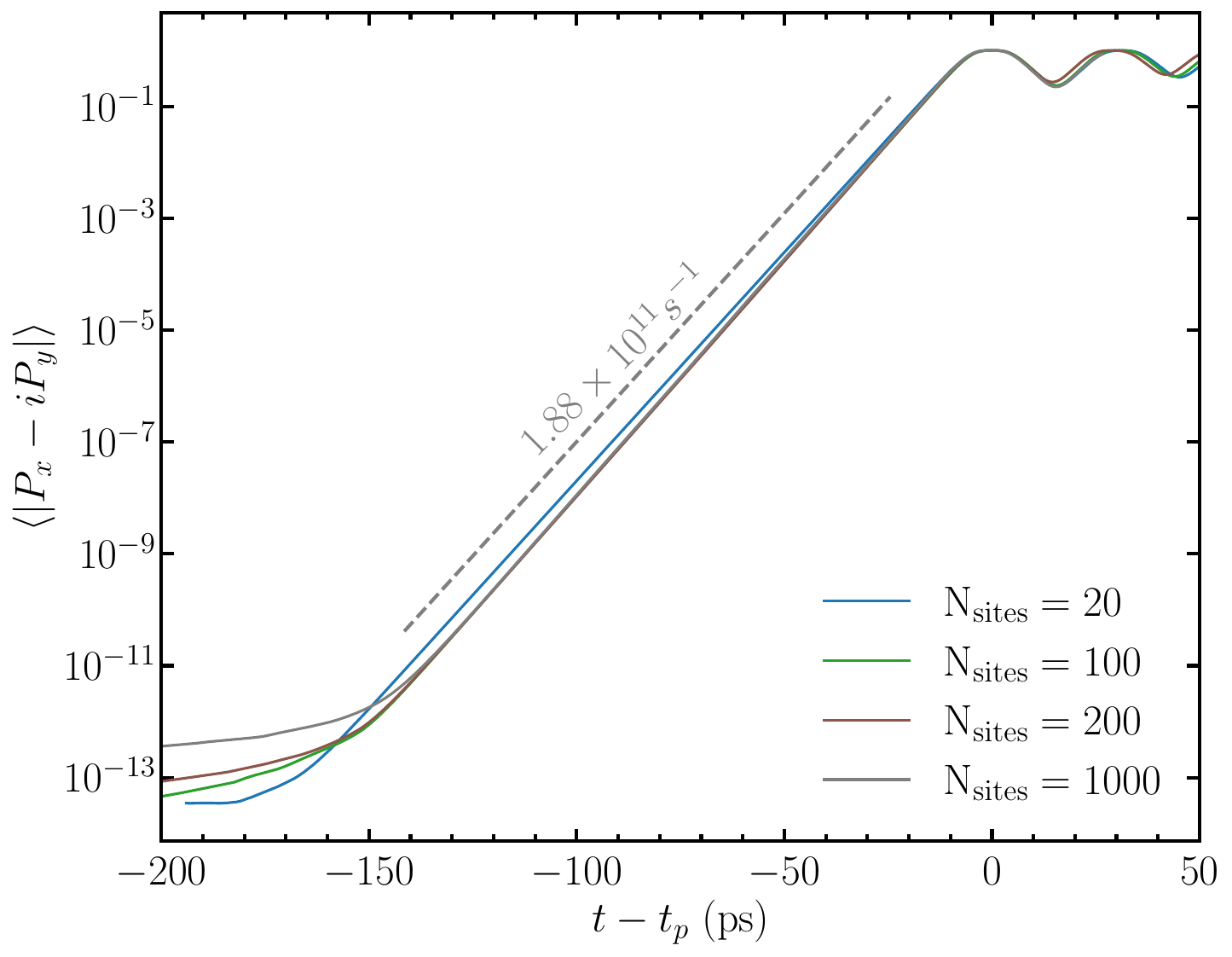}
\caption{Exponential growth of average single-site flavor coherence as a function of time and system size for the flavor-symmetric two-beam FFI. Curves are aligned at their first peak for visual comparison of the growth rates with the value obtained from linear stability analysis (dashed line).}
\label{fig:growth_rate_EMU_vs_CCNO_diff_N_w_error_domain_avg_periodic}
\end{figure}
We check resolution requirements for resolving the linear growth phase of the fast flavor instability and display the results in Figure~\ref{fig:growth_rate_EMU_vs_CCNO_diff_N_w_error_domain_avg_periodic}. Specifically, the 20-particle system achieves a growth rate of $1.88\times10^{11}\,\text{s}^{-1}$ with only $0.186\%$ error, while simulations with larger particle numbers yield slightly higher deviations: $1.93\times10^{11}\,\text{s}^{-1}$ for 100 particles ($2.61\%$ error), $1.928\times10^{11}\,\text{s}^{-1}$ for 200 particles ($2.537\%$ error), and $1.905\times10^{11}\,\text{s}^{-1}$ for 1000 particles ($1.151\%$ error). These numerical discrepancies primarily stem from finite timestep size, with variations depending on the chosen timestep $\tau$. Following this exponential phase, the growth saturates due to nonlinear effects and mode coupling, causing the averaged off-diagonal density amplitudes to saturate at a few tenths of the total neutrino density.

\begin{figure}
    \centering
        \includegraphics[width=1.0\linewidth]{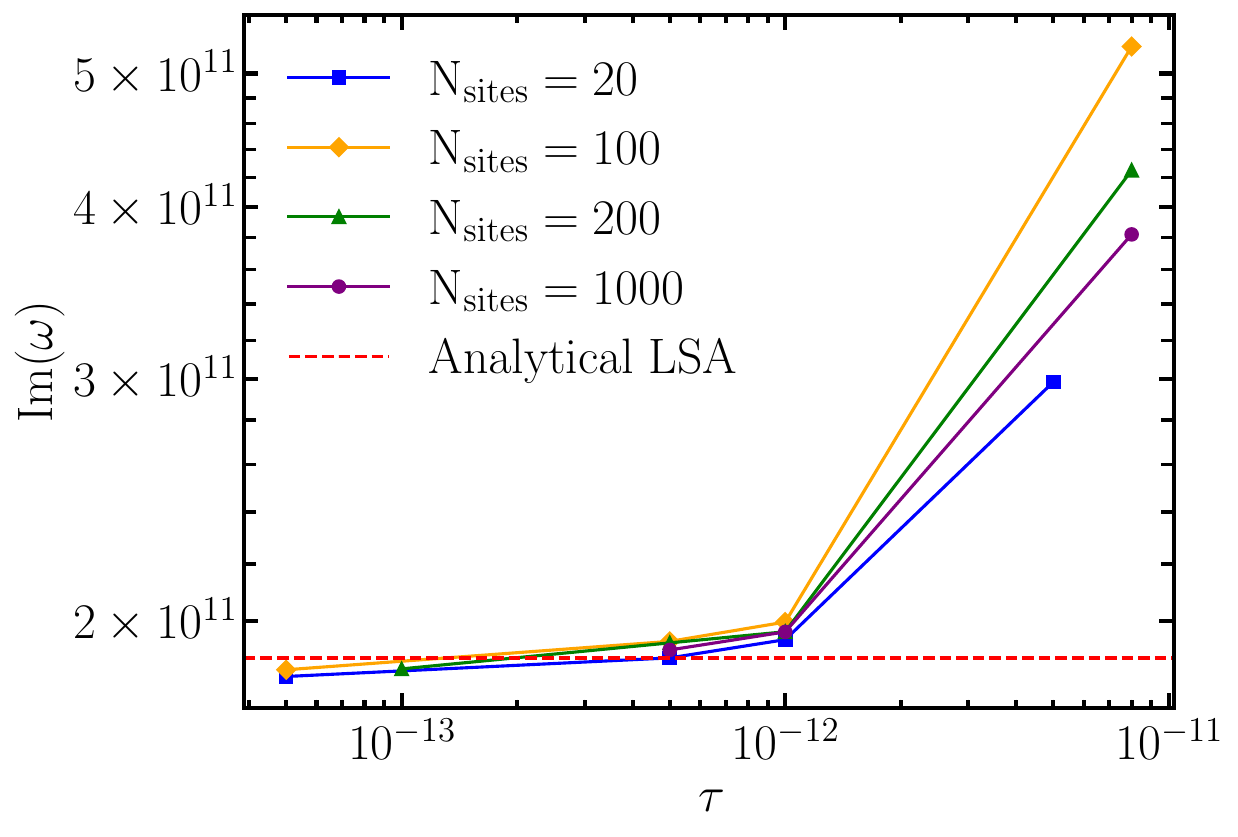}
        \caption{Mean-field inhomogeneous FFI growth rate $\mathrm{Im}(\omega)$ as a function of the timestep size $\tau$ for various system sizes $N_\mathrm{sites}$. Convergence is apparent with both the timestep $\tau$ and the number of particles. }
    \label{fig:time_step_vs_growth_rate_w_predictions}
\end{figure}
We conduct resolution tests as shown in Fig.~\ref{fig:time_step_vs_growth_rate_w_predictions}, analyzing the mean-field inhomogeneous FFI growth rates across various system sizes on a log-log scale, as functions of timestep size ($\tau$). The plot highlights clear asymptotic behavior: as the timestep is reduced and the number of particles is increased, the computed growth rates systematically approach the analytic linear stability prediction. At large timestep sizes ($\tau \gtrsim 10^{-12}\,\text{s}$), this convergence breaks down since each particle traverses distances greater than their spacing ($\Delta z/c$) within a single timestep. Hence, simulations with smaller $N_\mathrm{sites}$ display deviations earlier due to coarser spatial resolution. To establish an optimal balance between computational cost and numerical accuracy, we adopt a fiducial timestep of $\tau = 5\times10^{-13}\,\text{s}$ throughout the main results. This timestep ensures reliable results with less than $3\%$ error relative to analytic predictions across all studied system sizes for this bond dimension (BD = 1), capturing both the correct physical dynamics and numerical convergence effectively.

%========================%
% Asymmetric Convergence %
%========================%
\subsection{Asymmetric MF evolution and convergence behavior}\label{sec:results_IC_asymmetric}
\begin{figure}
    \centering
    \includegraphics[width=1.0\linewidth]{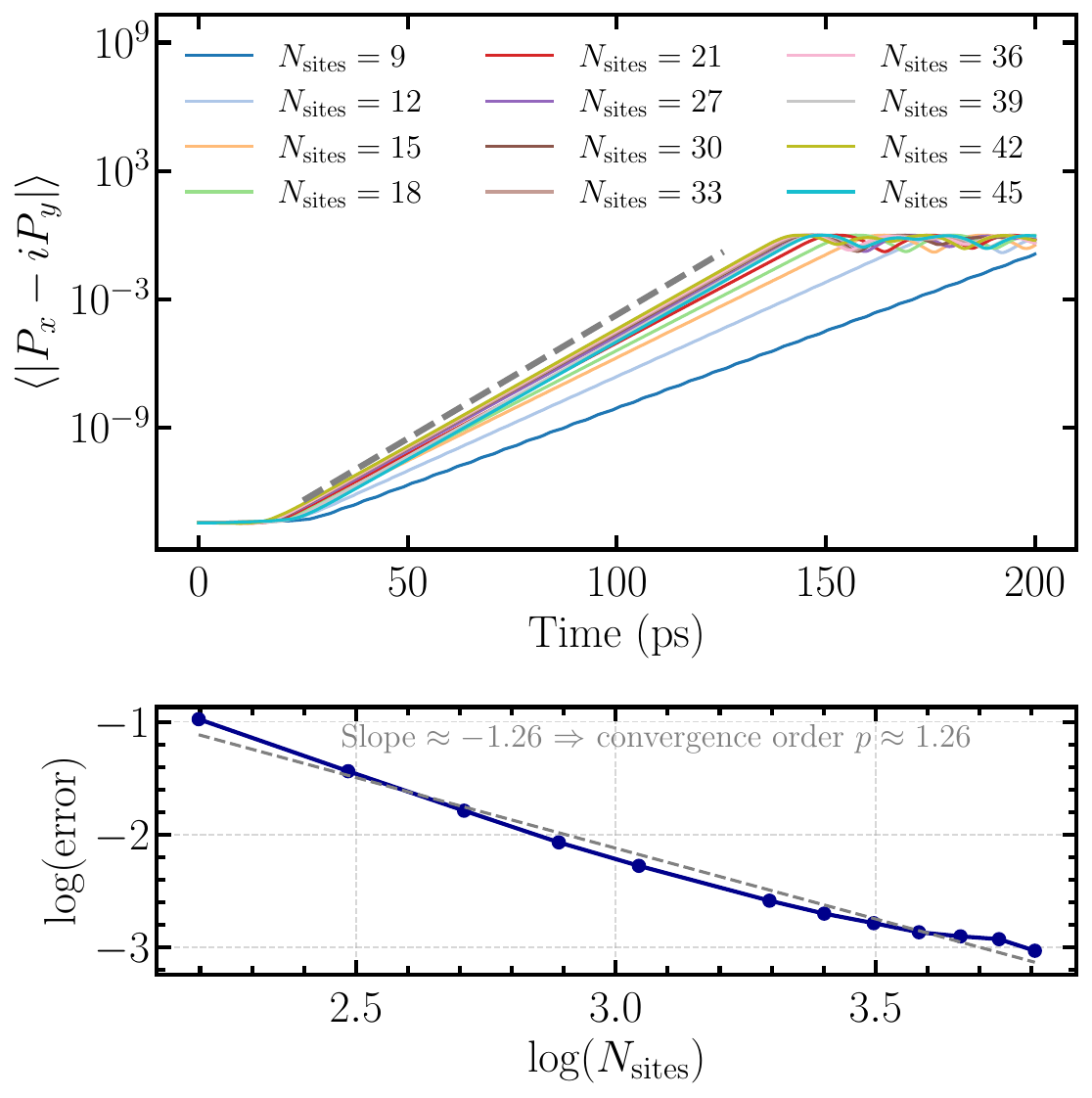}
    \caption{
    Comparison between numerical and analytical growth rates for asymmetric mean-field (MF) simulations. 
    \textbf{Top panel:} Time evolution of the domain-averaged flavor coherence for different total system sizes, $N_{\mathrm{sites}} = N_{\mathrm{sites}}^{\nu_\mu} + 2N_{\mathrm{sites}}^{\nu_e}$ (with $N_{\mathrm{sites}}^{\nu_\mu}=N_{\mathrm{sites}}^{\nu_e}$), compared against the analytical linear stability analysis (LSA) prediction, $\mathrm{Im}(\omega)=2.66\times10^{11}\,\mathrm{s^{-1}}$ (gray dashed line). The numerical growth rates converge to within $\sim5\%$ of the analytical result as spatial resolution increases. 
    \textbf{Bottom panel:} error as a function of $N_\mathrm{sites}$. The linear fit (gray dashed line) demonstrates a convergence order of $p\simeq1.26$.}
    \label{fig:rho_emu_vs_time_asymmetric_and_convergence}
\end{figure}
Building on the observed convergence of open-boundary systems toward closed-system MB behavior for superimposed configurations, we next re-assess the numerical consistency of our MF framework. In particular, we investigate whether the solver maintains the expected scaling with $N_{\rm sites}$ when applied to flavor-asymmetric initial conditions in a closed, superimposed setup. This test simultaneously benchmarks the MF solver against analytical linear stability predictions, ensuring that numerical resolution effects do not bias comparison with many-body results, and provides a physically meaningful reference for how flavor asymmetry evolves in the MF limit as system size increases.

Figure~\ref{fig:rho_emu_vs_time_asymmetric_and_convergence} (top panel) presents the temporal evolution of the asymmetric-flavor coherence $\langle|\rho_{e\mu}|\rangle$ for different total system sizes $N_{\mathrm{sites}} = N_{\mathrm{sites}}^{\nu_\mu} + 2N_{\mathrm{sites}}^{\nu_e}$ (with $N_{\mathrm{sites}}^{\nu_\mu}=N_{\mathrm{sites}}^{\nu_e}$). Across all resolutions, the exponential growth rate agrees closely with the analytical linear stability analysis (LSA) prediction $\mathrm{Im}(\omega)=2.66\times10^{11}\,\mathrm{s^{-1}}$, with deviations below $5\%$ for fine spatial resolutions. This confirms that the implemented MF solver reliably captures the physical instability growth even in asymmetric configurations.

To verify the robustness of these results, the bottom panel examines the numerical convergence behavior. The relative error from the analytical prediction, $\epsilon$, scales as
\(
\epsilon \propto N_{\mathrm{sites}}^{-p},
\)
where $N_{\mathrm{sites}}$ is the total number of discretized sites and $p$ is the convergence order. Taking logarithms gives 
  \(
  \log(\epsilon) = a - p\,\log(N_{\mathrm{sites}}),
\)
  where the slope of the log--log plot directly yields $-p$. The best-fit slope of $-1.26$ therefore corresponds to $p \approx 1.26$, indicating near first-order convergence. This confirms that the implemented discretization scheme exhibits consistent and predictable accuracy, with numerical errors diminishing approximately as $(\Delta z)^{1.26}$ as spatial resolution of particles improves.

% {plots/t_p_vs_symmetric del_omega.pdf}
%===========%
% Many-Body %
%===========%
\section{Many-body tests}
\begin{figure}
    \centering
        \includegraphics[width=\linewidth]{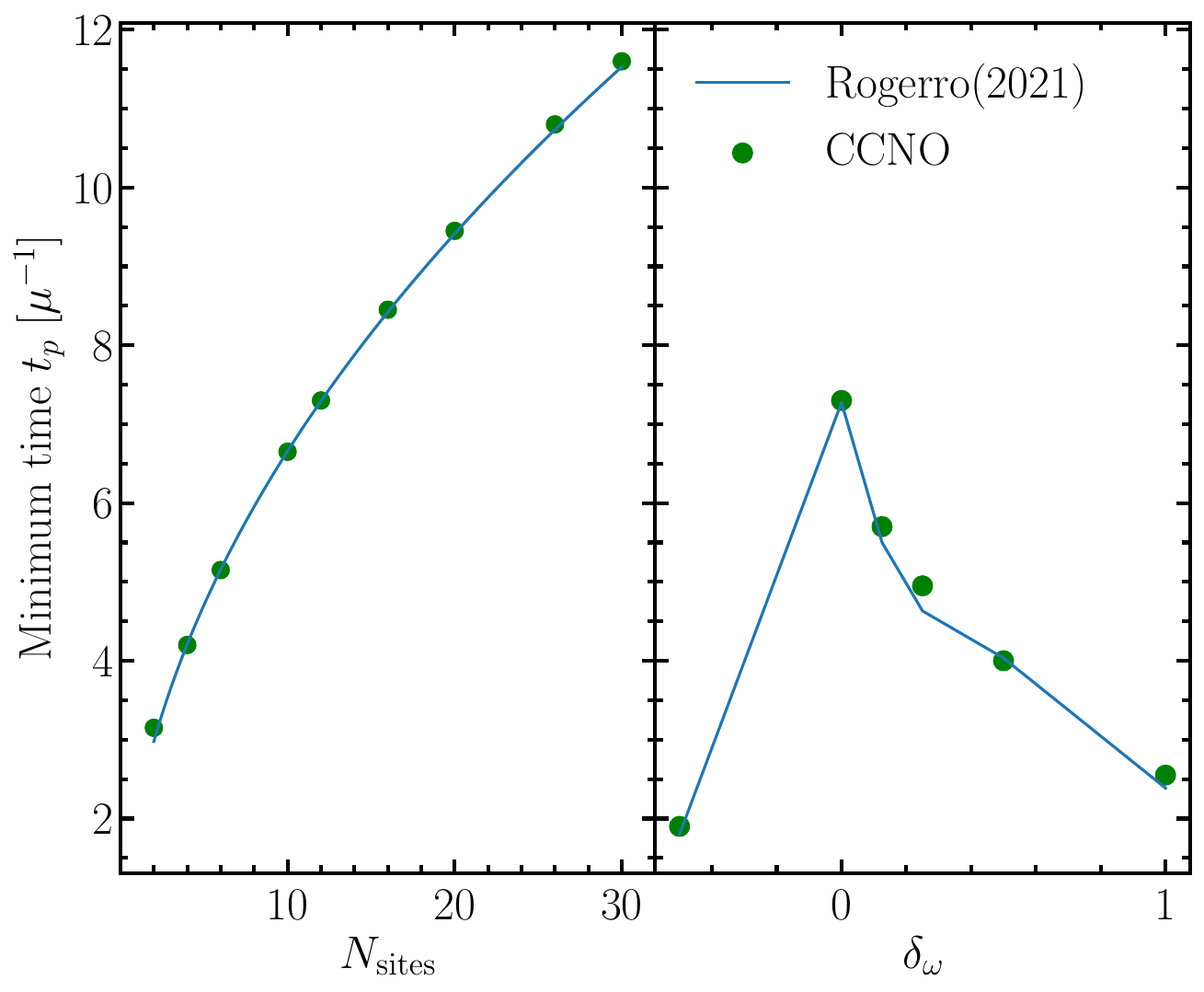}
    \caption{Variation with the $N_{\rm sites}$ and asymmetry energy $\delta_{\omega}$ plotted against the time taken to reach the minimum of the survival probability $t_P$. The blue curve plots the expected values as determined by the empirical fitting function given in \cite{roggero2021entanglement}.}
    \label{fig:Table.1 Rog}
\end{figure}
Ref.~\cite{roggero2021entanglement} calculated the time of the first minimum of the survival probability and fitted the results to functions of parameters $(\delta\omega/\mu)=(\omega_A-\omega_B)/2\mu$ and $N_\mathrm{sites}$ (see their Table~I), where half of the sites are placed in the $A$ group and half in the $B$ group. The following describes our choice of parameters to replicate their results, where all units of mass and energy are arbitrarily given units of ergs in the code. For simplicity, we set $N_i=\Delta z^3/\sqrt{2}G_F N_\mathrm{sites}$ such that $\mu=1$ (arbitrarily choosing $\Delta z=10^{-3}\,\mathrm{cm}$) and set mixing angle $\theta=0$. We set the shape function to $S(\xi)=1$ and the geometric function to $J_{ij}=1$. We set $E_{i\in A}=0.5$ and $E_{i\in B}=-0.5$. In cases where $(\delta\omega/\mu)<0$ we set $m_1=\sqrt{|\delta\omega/\mu|}$ and $m_2=0$, whereas in cases where $(\delta\omega/\mu)>0$ we set $m_1=0$ and $m_2=\sqrt{\delta\omega}$. These parameters produce the desired values of $\delta\omega/\mu$ without requiring us to change any source code.

We set the initial state to $ |\Psi_0\rangle = \bigotimes_{n=1}^{(N_\mathrm{sites}/2) \in A} |\downarrow \rangle \otimes \bigotimes_{m=1}^{(N_\mathrm{sites}/2) \in B} |\uparrow \rangle$. We simulate for a total time of $15\,\mathrm{erg}^{-1}\times \hslash$ using time steps of size $0.05 \,\mathrm{erg}^{-1}\times \hslash$. For all tests, we set the maximum bond dimension to 1000, which implies no truncation for any of the calculations in this appendix.

In the left panel of Fig.~\ref{fig:Table.1 Rog} we show the time of the first minimum in the survival probability for the case of $\delta\omega/\mu=0$. The fit from \cite{roggero2021entanglement} is plotted in blue, with our results plotted as green dots. We show good agreement with their results describing the behavior of the system with increasing system size.

We also show the same quantities under variations of $\delta\omega/\mu$ under a constant system size of $N_\mathrm{sites}=12$ in the right panel of Fig.~\ref{fig:Table.1 Rog}. Once again, our method is able to reproduce the patterns resulting from the relative size of the vacuum and self-interaction potentials.

\end{document}